\documentclass[aps,showpacs,prb,superscriptaddress,floatfix,twocolumn,citeautoscript]{revtex4-1}
\usepackage{amstext}
\usepackage{graphicx}
\usepackage{amsmath}
\usepackage{amssymb}
\usepackage{bm}
\usepackage{dcolumn}
\usepackage{subfigure}
\usepackage{units}
\usepackage{hyperref}
\usepackage[usenames]{color}
\usepackage{booktabs}
\usepackage{multirow}
\usepackage{siunitx}
\usepackage{wrapfig}
\usepackage{lipsum}
\usepackage{setspace}
\hypersetup{pdfborder=0 0  0,colorlinks=true,citecolor=blue,linkcolor=blue}
\input epsf
\newcommand{\BN}{\textit{h}-BN}

\newcommand{\MS}{MoS$_2$}

\begin{document}
	
	\title{High temperature electron-hole superfluidity with strong anisotropic gaps in double phosphorene monolayers}
	
		\author{S. Saberi-Pouya}
		\email{S\_Saberi@sbu.ac.ir}
		\affiliation{Department of Physics, Shahid Beheshti University, G. C., Evin, Tehran 1983969411, Iran}
		\affiliation{Department of Physics, University of Antwerp, Groenenborgerlaan 171, B-2020 Antwerpen, Belgium}
		\author{M. Zarenia}
		\email{
			mohammad.zarenia@uantwerpen.be}
		\affiliation{Department of Physics, University of Antwerp, Groenenborgerlaan 171, B-2020 Antwerpen, Belgium}
		\affiliation{Department of Physics and Astronomy, University of Missouri, Columbia, Missouri 65211, USA}
		\author{A. Perali}
		\affiliation{School of Pharmacy, Physics Unit, University of Camerino, 62032 Camerino (MC), Italy}
		\author{T. Vazifehshenas}
		\affiliation{Department of Physics, Shahid Beheshti University, G. C., Evin, Tehran 1983969411, Iran}
		\author{F. M. Peeters}
		\affiliation{Department of Physics, University of Antwerp, Groenenborgerlaan 171, B-2020 Antwerpen, Belgium}
	\begin{abstract}
		
		 Excitonic superfluidity in double phosphorene monolayers is investigated using the BCS mean-field equations. Highly anisotropic superfluidity is predicted where we found that the maximum superfluid gap is in the BEC regime along the armchair direction and in the BCS-BEC crossover regime along the zigzag direction. We estimate the highest Kosterlitz-Thouless transition temperature with maximum value up to $\sim 90$ K with onset carrier densities as high as $4 \times 10^{12}$ cm$^{-2}$. This transition temperature is significantly larger than what is found in double electron-hole few-layers of graphene. Our results can guide experimental research towards the realization of anisotropic condensate states in electron-hole phosphorene monolayers.
		
	\end{abstract}
	\date{\today}
	\pacs{71.35.-y, 73.21.-b, 74.78.Fk}
	\maketitle
	
	\section{Introduction}
	
Black phosphorus (BP) is the most thermodynamically stable phase of phosphorus at ambient temperature and pressure \cite{PhysRev.92.580,John:sc63}. BP consists of puckered hexagonal layers coupled through the weakly van der Waals interlayer interactions \cite{PhysRevB.86.035105,Likai:nat17}. Such as graphene and hexagonal boron nitride (h-BN), monolayer and few-layer of BP can be exfoliated from its bulk material\cite{Xia:nat13}. Phosphorene, a monolayer of BP, has a direct energy gap \cite{Neto:rev2016,PhysRevB.89.235319} and exhibits high carrier mobility \cite{liu:acs14} which has recently attracted significant attention as a new 2D semiconductor material for electronic and optical applications\cite{liu:acs14,PhysRevB.93.075408,PhysRevLett.114.066803,PhysRevB.90.205421,PhysRevB.94.085417}. A special feature of phosphorene is the high in-plane anisotropy of its energy band structure. This anisotropy comes from the layered puckered-honeycomb structure of phosphorene resulting from its \textit{sp$^3$} hybridization. Motivated by this peculiar property, several theoretical and experimental studies investigated different anisotropic properties of phosphorene \cite{Xiaomu:nat15,Samira:Conductivity2017,qiao2014high}. Recently, many-body aspects of phosphorene have been also addressed through the study of collective excitation modes \cite{2053-1583-4-2-025064,Samira:SOphonon2017,Low:prl14,Jin:prb15,Rodin:prb15} in a doped monolayer and Coulomb drag\cite{Samira:drag2016} in coupled phosphorene sheets.

Phosphorene is unstable in air \cite{2053-1583-2-1-011002,Favron:nat15} and therefore encapsulation with h-BN is used resulting in devices which are conductive and fully stable under ambient conditions\cite{Cao:nanol15,Likai:nat17}. There has been an increasing interest in the study of van der Waals heterostructures including 2D conducting sheets separated by thin h-BN insulating layers. The interest is mainly because these systems offer the possibility for the observation of a coherent superfluid state in spatially separated electron- and hole-doped conducting sheets driven by the strong inter-layer Coulomb interaction. Unlike conventional double-GaAs quantum well structures, which typically have inter-layer electron-hole separation larger than the effective Bohr radius,
the separation between the electron-hole sheets in 2D van der Waals heterostructures \cite{Gorbachev:natphys12}, can be as small as 1 nm (i.e. three h-BN layers) and still provide a potential barrier high enough to eliminate inter-layer tunneling.
This leads to a strong inter-layer coupling, boosting the onset of superfluidity. 
 The anisotropic superfluidity for quasiparticles was studied in first time for He$^3$\cite{PhysRevLett.31.870}. Following this, a very substantial fraction of theoretical\cite{PhysRevB.57.R6846,PhysRevB.78.140502,PhysRevB.89.060502,Zareniasuperfluidity:2014} and experimental\cite{PhysRevLett.101.246801, PhysRevLett.117.046803,Liu:nat17} work has been devoted to the study of superfluidity.
Electron-hole superfluidity has been studied in different double graphene systems, including monolayers \cite{PhysRevB.86.045429}, bilayers \cite{PhysRevLett.110.146803}, few layers\cite{Zareniasuperfluidity:2014}, nanoribbons \cite{Zarenia:nanoribbon16} and hybrid graphene-GaAs quantum heterostructures \cite{PhysRevLett.111.086804}. Such systems below the Kosterlitz-Thouless (KT) transition temperature may support a superflow of electron-hole pairs. Except for the double electron-hole monolayer graphene system, where strong screening kills any superfluidity \cite{PhysRevB.86.045429}, the other graphene systems show promising predictions for the observation of a condensate superfluid state. Indeed, double bilayer\cite{PhysRevLett.110.146803} or few layer graphene \cite{Zareniasuperfluidity:2014} systems can access the strong electron-hole pairing regime at low enough densities, where large superfluid gaps are able to suppress the detrimental Coulomb screening, even before the bosonic limit of point-like excitons is reached. The high-temperature superfluidity of two-dimensional dipolar excitons in two parallel transition metal dichalcogenide (TMDs) layers was also predicted\cite{PhysRevB.93.245410}. In addition it appears that the bilayer graphene heterostructures, a system which has been theoretically predicted to support a stable exciton superfluid  demonstrate remarkable progresses in excitonic superfluidity as well as Coulomb drag experiments\cite{PhysRevB.95.045416,PhysRevLett.110.146803,PhysRevLett.117.046803,PhysRevLett.117.046802}. Electron-hole pair condensation has been measured in a graphene/\MS\ heterostructure below 10K, in the absence of an external magnetic field\cite{Joo2017}. Very recently, it was shown that the strongly enhanced tunneling between bilayer graphene sheets separated by bilayer WSe$_2$ can be experimentally observed at zero magnetic field and at the neutrality condition\cite{GrapheneWSe2}.

\begin{figure}[t]
	\includegraphics[width=9.cm]{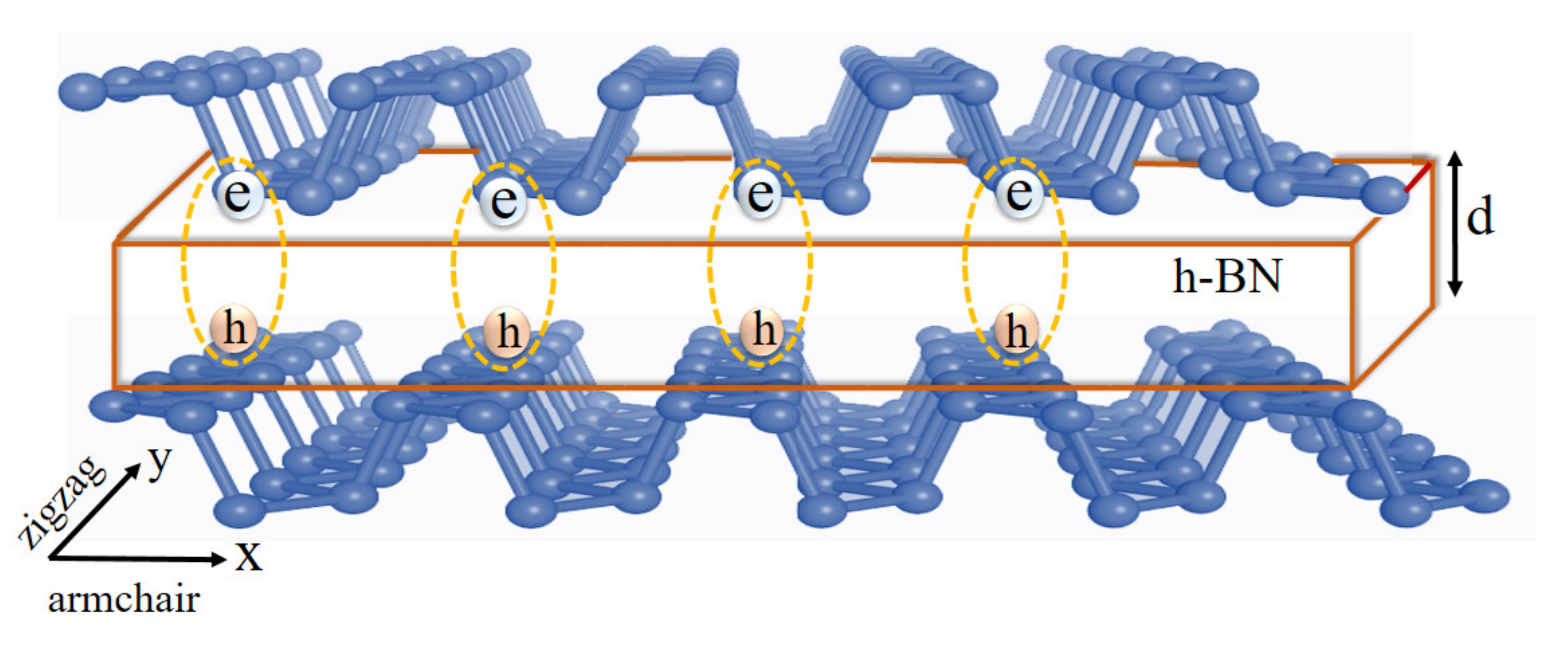}
	\caption{Schematic illustration of two phosphorene sheets separated by a thin barrier of h-BN layers. The electrons and holes are induced by top and back gates in the separately electrically contacted upper and lower phosphorene sheets. }
	\label{fig:1}
\end{figure}	

In this paper, we investigate the possibility of \emph{anisotropic superfluidity} in double electron-hole phosphorene sheets. We found that a highly anisotropic superfluid state occurs in double phosphorene sheets arising from the anisotropic low energy bands in phosphorene. Because of the anisotropic charge carrier effective mass along the armchair and zigzag directions, excitons in phosphorene are characterized by a strong spatial anisotropy and can exist at elevated temperatures with large binding energies\cite{Xiaomu:nat15,PhysRevB.93.121405}.
Our system consists of two parallel monolayer phosphorene sheets (see Fig. \ref{fig:1}). $X$ and $y$ directions correspond to the armchair and zigzag directions of phosphorene lattice, respectively. The upper sheet of electrons and the lower sheet of holes are controlled by the top and back gates. The two sheets are separated by a thin h-BN insulating barrier to prevent tunneling between the sheets and electron-hole recombination. The semiconductor nature of phosphorene with a large energy gap of $\sim1.5-2$ eV allows us to restrict ou calculations to the conduction band. We first extend the isotropic mean-field equations for the superfluid gap and superfluid density to the case of anisotropic energy bands considering static screened electron-hole Coulomb interaction. We calculate the superfluid energy gap and transition temperature to see whether a superfluid state can form in the double phosphorene system in experimentally attainable densities. We predict a highly anisotropic gap function that depends on the direction of motion of the electron-hole pair, caused by the anisotropic band structure. Using Kosterlitz-Thouless relation, we also estimate an upper limit for the transition temperature. The approach outlined in this paper is general and will be easily adaptable to other emerging anisotropic 2D materials. Our numerical analytics will be applied to phosphorene. Moreover, an analytical approach for indirect excitons in a phosphorene double layer was studied in Ref. [\onlinecite{PhysRevB.96.014505}]. Suggested in Ref. [\onlinecite{PhysRevB.96.014505}] for an electron-hole double layer of black phosphorus,  the analytical expressions for the single dipolar exciton energy spectrum and wave function was obtained. It was predicted that a weakly interacting gas of dipolar excitons in a double layer of black phosphorus exhibits superfluidity due to the dipole-dipole repulsion between the dipolar excitons.  The angle-dependent sound velocity was calculated, which causes the dependence of the critical velocity for the superfluidity on the direction of motion of dipolar excitons.  In the present article, we extend the isotropic mean-field equations for the superfluid gap and superfluid density to the case of anisotropic energy bands considering static screened electron-hole Coulomb interaction. We employ a mean field approach in which the finite energy gap in the excitation spectrum of the system $\Delta(\mathbf{k})$ will be calculated in the presence of an effective screened electron-hole pairing interaction. Our results obtained with this model indicate, in contrast to Ref. [\onlinecite{PhysRevB.96.014505}], that we are in the BEC regime of strong coupling along the armchair direction and in the BEC-BCS crossover regime of weaker coupling along the zigzag direction. Remarkably, due to the strong screening effect the superfluid gap vanishes in the weakly interacting BCS regime. 

The paper is organized as follows: In Sec. \ref{theory} we present 
 the mean-field equations of the superfluid gap and density as well as the polarization functions of the screened Coulomb interaction and then extend them for a system with anisotropic single-particle energy bands. Sec. \ref{result} deals with the anisotropic gap function for double-layer phosphorene and implement numerically the self-consistent method to phosphorene. Finally, we conclude in Sec. \ref{conclude}.
	
	\section{theory}
	\label{theory}
	\subsection{Isotropic mean field equation}
	
	We employ a mean field approach in which the finite energy gap in the excitation spectrum of the system $\Delta(\boldsymbol{k})$, is a signature of the superfluid ground state. For a system with equal density of electrons and holes $n_{e}=n_{h}=n$, the pair excitation energy and the equation for the momentum dependent gap function at zero temperature are, respectively\cite{GORTEL1996146}
	
	\begin{equation}
		E(\boldsymbol{k})=\sqrt{\xi^2(\boldsymbol{k})+\Delta^2(\boldsymbol{k})},
		\label{eq:5}
	\end{equation}
	
	\begin{equation}
		\Delta(\boldsymbol{k})=-\frac{1}{\Omega}\sum_{\boldsymbol{q}} V(q) \frac{\Delta(\boldsymbol{k}-\boldsymbol{q})}{2E(\boldsymbol{k}-\boldsymbol{q})},
		\label{eq:6}
	\end{equation}
	
	\noindent and the mean field equation for the electron(hole) density is given by
	
	\begin{equation}
		n=\frac{g_s g_\nu}{\Omega}\sum_{\boldsymbol{k}^{\prime}} \frac{1}{2} \bigg(1-\frac{\xi(\boldsymbol{k}^{\prime})}{E(\boldsymbol{k}^{\prime})}\bigg),
		\label{eq:7}
	\end{equation}
	
	\noindent where $\Omega$
	is the surface area of the system, $\xi(\boldsymbol{k})=[\xi^{e}(\boldsymbol{k})+\xi^{h}(\boldsymbol{k})]/2$ with $\xi^{e/h}(\boldsymbol{k})$ the energies of electrons and holes
	measured from the chemical potential $\mu$ and $g_s=2$ is the spin degeneracy.  In contrast to graphene, there is no valley degeneracy in phosphorene \cite{Rodin:prl14,qiao2014high} ($g_\nu=1$), which leads to a weaker Coulomb screening as compared to graphene, as will be shown below. The screened Coulomb interaction , $V(q)$, between electron(e) and hole(h) layers in the random phase approximation (RPA) is given by \cite{PhysRevB.85.195136}

		\begin{equation}
		V(q)=\frac{v_{d}+\Pi_a (v^2_q-v_d^2)}{1+2(v_q \Pi_n+v_d \Pi_a)+(v_q^2-v_d^2)(\Pi_n^2-\Pi_a^2)},
		\label{eq:8}
		\end{equation}

	\noindent Here $v_{q}= -2\pi e^{2}/\kappa q$ and $v_{d}= v_q \exp({-qd})$ are respectively, the intra-layer and inter-layer Coulomb interaction screened by a surrounding medium with dielectric permitivity $\kappa$. Here we set $\kappa = 3$ which is the dielectric constant of h-BN. $\Pi_{n}$ and $\Pi_{a}$ are the normal and anomalous polarization, respectively, and within the RPA are given by\cite{Hwang:prb09,PhysRevB.87.085401}
	
	\begin{equation}		
		\Pi_{n}(\boldsymbol{q})=-g_{s} \sum_{\boldsymbol{k}} \frac{u^2(\boldsymbol{k})
			\nu^2(\boldsymbol{k}-\boldsymbol{q})+\nu^2(\boldsymbol{k})\ u^2(\boldsymbol{k}-\boldsymbol{q})}{E(\boldsymbol{k})+E(\boldsymbol{k}-\boldsymbol{q})},
		\label{eq:10}
	\end{equation}
	
	\begin{equation}		
		\Pi_{a}(\boldsymbol{q})=g_{s} \sum_{\boldsymbol{k}} \frac{2u(\boldsymbol{k}) \nu(\boldsymbol{k}-\boldsymbol{q})  \nu(\boldsymbol{k}) \ u(\boldsymbol{k}-\boldsymbol{q})}{E(\boldsymbol{k})+E(\boldsymbol{k}-\boldsymbol{q})},
		\label{eq:11}
	\end{equation}
	
	\noindent where $u(\boldsymbol{k})$ and $\nu(\boldsymbol{k})$  are the coherence factors given by
	
	\begin{equation}
		u^{2}(\boldsymbol{k})=\frac{1}{2}\bigg(1+\frac{\xi(\boldsymbol{k})}{E(\boldsymbol{k})}\bigg) , \	
		\ \nu^{2}(\boldsymbol{k})=\frac{1}{2}\bigg(1-\frac{\xi(\boldsymbol{k})}{E(\boldsymbol{k})}\bigg). \\
		\label{eq:12}
	\end{equation}
	
	\noindent We note that $u(\boldsymbol{k}) \nu(\boldsymbol{k})=\Delta(\boldsymbol{k})/2E(\boldsymbol{k})$. The most favorable conditions for pairing occurs when $k_F d \ll 1$. In this limit,  Eq. (\ref{eq:8}) can be approximated as\cite{PhysRevLett.110.146803,Zareniasuperfluidity:2014}
		
		\begin{equation}
			V(q) \approx \frac{v_{q} \exp(-qd)}{1+2v_{q}(\Pi_{n}[\boldsymbol{q})+\Pi_{a}(\boldsymbol{q})]}.
			\label{eq:epsilon}
		\end{equation}
		
	\begin{figure*}[ht]
		\includegraphics[width=18.2cm]{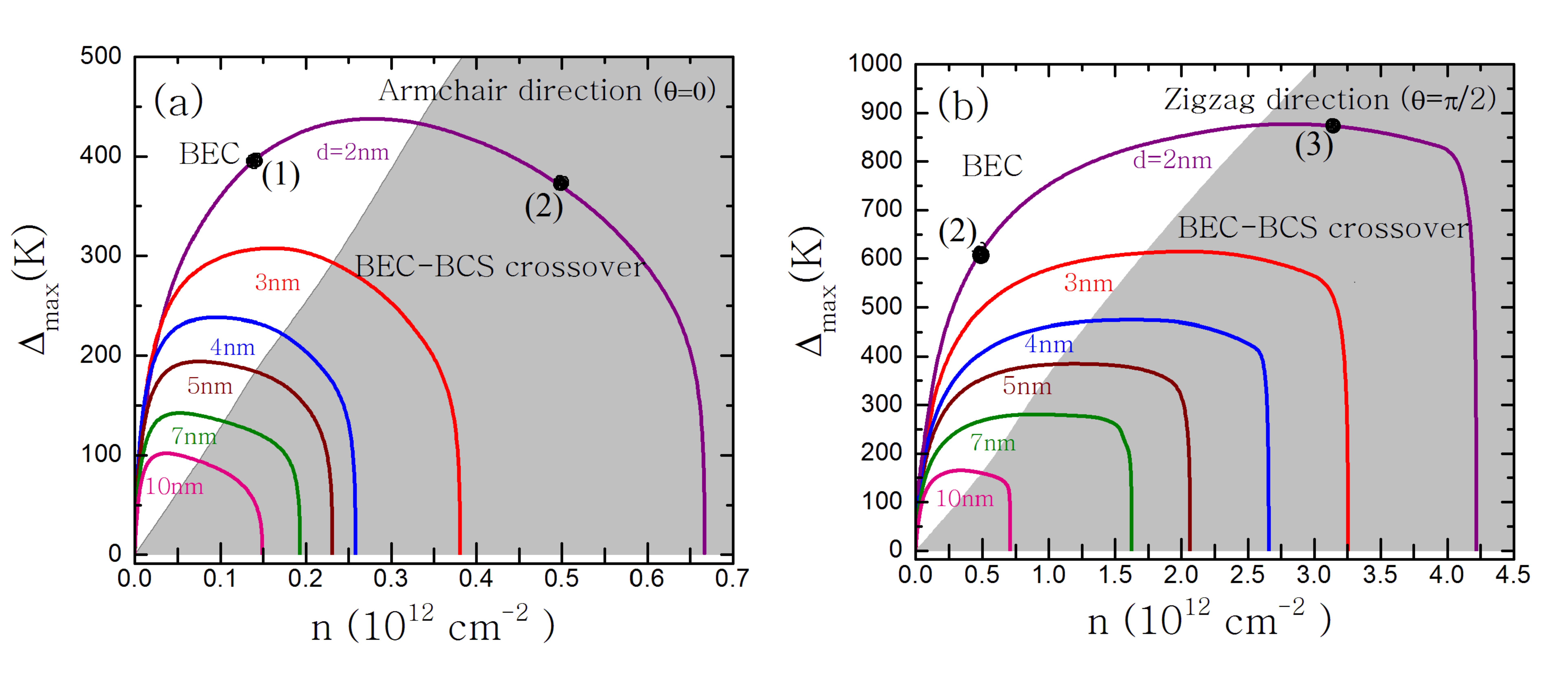}
		\caption{Maximum of superfluid gap along two main crystallographic directions of phosphorene: (a) the armchair direction ($\theta=0$) and (b) the zigzag direction ($\theta=\pi/2$) for different h-BN effective barrier thickness $d$. Shadow grey area indicates the intermediate BEC-BCS crossover regime for pairing.}
		\label{Fig:2}
	\end{figure*}
		
	\subsection{Anisotropic mean field equations}
	
	 Here, we extend the mean field equations (\ref{eq:6}) and (\ref{eq:7}) to the case of e-h phosphorene sheets where the energy bands are anisotropic. The corresponding hamiltonian of BP near the $\Gamma$ point can be expressed as \cite{Rodin:prl14}
	
	\begin{equation}
		\hat{H}_{\textbf{0}}=\left(\begin{array}{cc}
			E_c + \eta_c k^2_x + \alpha_c k^2_y 
			& \gamma k_x+\beta k_y^2\\
			\gamma k_x+\beta k_y^2 &E_v-\eta_v k^2_x - \alpha_v k^2_y 
		\end{array}\right),
		\label{eqH}
	\end{equation}
	
	\noindent where $x$ and $y$ stand for the armchair and zigzag directions, respectively (see Fig. \ref{fig:1}). $E_{c}$ ($E_{v}$) is the energy of conduction (valence) band edge,
	$\gamma$ and $\beta$ describes the effective coupling between the conduction and valence bands. $\eta_{c,v}$ and $\alpha_{c,v}$ are related to the known anisotropic effective masses of phosphorene\cite{Rodin:prl14}
		
	 \begin{equation}
	\begin{aligned}
	&m^x_{e}=\dfrac{{\hbar^2}}{2(\eta_c+\gamma^2/2E_g)},\\
	&m^x_{h}=\dfrac{{\hbar^2}}{2(\eta_v-\gamma^2/2E_g)},\\
	&m^y_{e(h)}=\dfrac{{\hbar^2}}{2(\alpha_{c(v)})},\\
	\end{aligned}\label{16}
	\end{equation}

\noindent where $E_g$ is the energy band gap. One can then use these masses to obtain an approximation for the spectrum \cite{Low:prl14,PhysRevB.93.165402}

	\begin{equation}
		\xi^{e(h)}(\boldsymbol{k})=\frac{\hbar^2}{2}\bigg(\frac{k_{x}^{2}}{m_{x}^{e(h)}}+\frac{k_{y}^{2}}{m_{y}^{e(h)}}\bigg)-\mu^{e(h)},
		\label{eq:1}
	\end{equation}
				 	
\noindent where $m^{e/h}_{x}$ and $m^{e/h}_{y}$ are the electron/hole effective masses along zigzag ($x$) and armchair ($y$) direction in each layer. Here we use the effective masses $m_{x}^{e}=m_{x}^{h}\approx 0.15m_{0}$ and $m_{y}^{e}\approx 0.7m_{0}$ and $m_{y}^{h}\approx 1.0 m_{0}$ ($m_{0}$ is the free electron mass)\cite{Low:prb14}. Because phosphorene is a semiconductor with a direct band gap at the $\Gamma$ point of the first Brillouin zone \cite{PhysRevB.89.235319}, inter-band transitions require extremely large energies. A detailed investigation of multiband effects on e-h superfluidity as function of the band gap energy is reported in Ref. [\onlinecite{PhysRevLett.119.257002}]. Here, we include only one band and limited ourselves to contributions from the conduction band of each layer. In this case, we use equal effective masses for the  $e$ and $h$ in each phosphorene sheet, so that we may drop the $e$ and $h$ indices. By using the anisotropic energy band Eq. (\ref{eq:1}),  the well-known analytical form of the normal polarization Eq.(\ref{eq:10}) and the anomalous polarization Eq.(\ref{eq:11}) will also be direction dependent in $q$ space. Here, to get rid of the anisotropy, we use the following transformation \cite{Low:prl14}

\begin{equation}
 \boldsymbol{s}=\sqrt{m_{D}/M}\boldsymbol{q}  \ \ \ \text{and} \ \ \  \boldsymbol{p}=\sqrt{m_{D}/M}\boldsymbol{k},
 \label{eq:trans}
 \end{equation}
 
 \noindent where $M$ is the mass tensor with diagonal elements $m_{x}$ and $m_{y}$, and $m_{D}=\sqrt{m_{x}m_{y}}$ is the 2D density of state mass. We consider $\boldsymbol{k}=k(\cos(\theta),\sin(\theta))$ where $\theta$ is the polar angle of the $\boldsymbol{k}$ vector with respect to the $x$ axis. So, we can rewrite  $p(\theta)=k\sqrt{m_{D}R(\theta)}$  with the orientation factor $R(\theta)=\cos^{2}(\theta)/m_{x} + \sin^{2}(\theta)/ m_{y}$ \cite{Rodin:prb15}. Using this simplified notation, the anisotropic energy $\xi(\boldsymbol{k})$ term becomes:
	
	\begin{equation}
			\xi(\boldsymbol{p})=\frac{\hbar^2 p^{2}}{2 m_D}-\mu.
		\label{eq:4}
	\end{equation}
		
\noindent Having this quadratic energy dispersion, we generalize the well-known analytical form of the polarizability to include anisotropy in the superfluid gap. Therefore, the zero temperature total polarization function can be calculated by using the renormalized isotropic energy bands Eq. (\ref{eq:4}) in Eq. (\ref{eq:10}) for the normal polarization function and in Eq. (\ref{eq:11}) for the anomalous polarization function.
	
	Using the polar notation, one can readily obtain the superfluid gap equation as
	
	\begin{equation}
		\Delta(\boldsymbol{p})=-\frac{1}{\Omega}\sum_{\boldsymbol{p^{\prime}}} V(\boldsymbol{p-p^{\prime}}) \frac{\Delta(\boldsymbol{p^{\prime}})}{2E(\boldsymbol{p^{\prime}})},
		\label{eq:16}
	\end{equation}
	
	\noindent and the density equation
	
	\begin{equation}
		n=\frac{g_s}{\Omega}\sum_{\boldsymbol{p^{\prime}}} \frac{1}{2} \bigg(1-\frac{\xi({\boldsymbol{p^{\prime}}})}{E({\boldsymbol{p^{\prime}}})}\bigg).
		\label{eq:19}
	\end{equation}
	
	Here we solve Eqs. (\ref{eq:16})-(\ref{eq:19}) self-consistently for the direction-dependent $\Delta({\boldsymbol{k}})$. It should be noted that while the spin orbit coupling (SOC) mainly affects the superfluid stiffness\cite{PhysRevA.88.063637,PhysRevA.91.023609}, one can ignore this effect on the superfluidity of phosphorene. It was shown that the SOC has no effect on the band structure of phosphorene\cite{PhysRevB.90.115439}.
	
		\begin{figure*}[ht]
			\includegraphics[width=18.5cm]{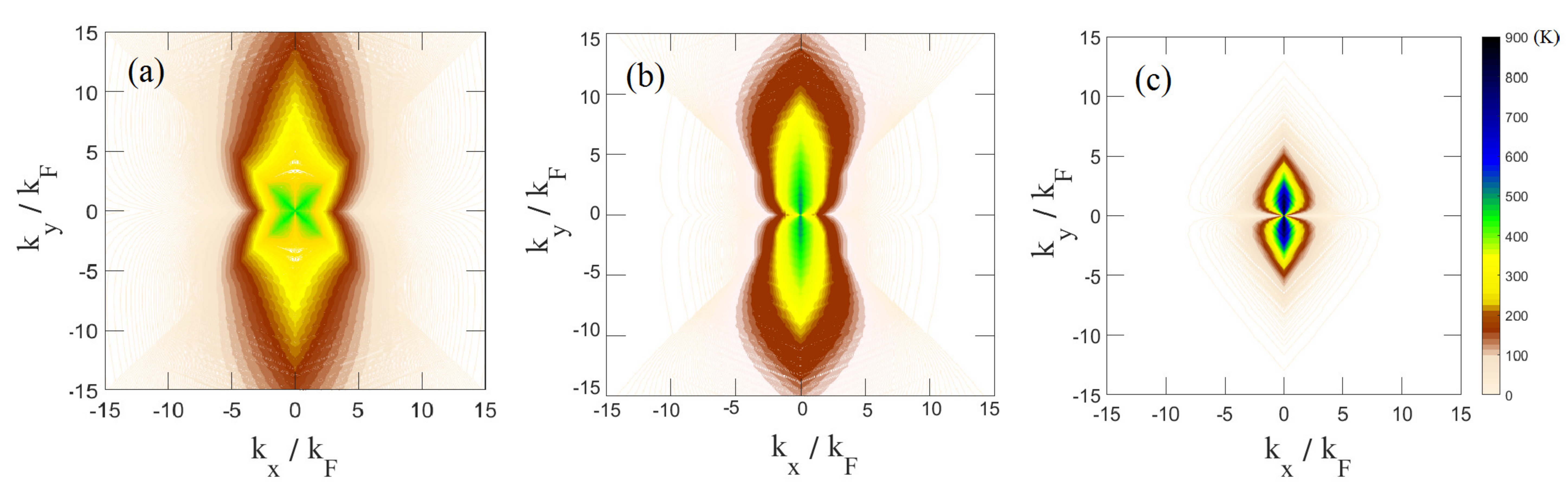}
			\caption{ Contour plot of $\Delta({\boldsymbol{k}})$ for density: (a) $0.15\times 10^{12} cm^{-2}$ , (b) $0.5\times 10^{12} cm^{-2}$, and (c) $3.25\times 10^{12} cm^{-2}$ which are marked in Fig. \ref{Fig:2}. Here the effective h-BN barrier thickness is $d=2\ \text{nm}$.}
			\label{Fig:3}
		\end{figure*}

	\section{Results and discussions}
	\label{result} 	
	
	In this section, we present and discuss our numerical
	results for superfluidity of e-h phosphorene monolayers. Fig. \ref{Fig:2} shows the maximum of superfluid gap $\Delta_{\text{max}}$ as a function of density for different interlayer separations $d$ and for both (a) armchair ($\theta=0$) and (b) zigzag ($\theta=\pi/2$) directions of phosphorene. With decreasing inter-layer spacing $d$ the pairing interaction in the two directions become stronger and superfluidity persists up to, $\text{e.g.} \ n\approx 4 \times 10^{12} \text{cm}^{-2} $ in zigzag direction for $d=2\ \text{nm}$. For densities above an onset density, $n>n_c$, the superfluid gap vanishes. It is the screening that kills the superfluidity before it reaches the BCS regime. The interplay between the BEC-BCS crossover and the superfluid screening of the e-h pairing interactions have been investigated in Ref. [\onlinecite{PhysRevB.89.060502}] and by Quantum Monte Carlo simulations\cite{Pablo2017}. We find that $\Delta_{\text{max}}$ is about twice larger along the zigzag direction than along the armchair direction for the same $d$.
	This effect can be linked to the larger effective mass of carriers along the zigzag direction resulting in a significant enhanced exciton e-h binding energy. 
 Therefore, in the zigzag direction the interactions are stronger than in the armchair direction enhancing the superfluidity.
 The BEC and BEC-BCS crossover regimes in Fig. \ref{Fig:2}
 are distinguished, respectively, by the white and grey regions. We find that the peak in $\Delta_{\text{max}}$ is located in BEC regime for both armchair and zigzag directions. The BEC and BEC-BCS crossover regimes in Fig. \ref{Fig:2} are determined by evaluating the condensate fraction,  $C_F=\sum_{\boldsymbol{k}}^{} u^2(\boldsymbol{k}) v^2(\boldsymbol{k})/\sum_{\boldsymbol{k}} v^2(\boldsymbol{k})$; $C_F>0.8$ corresponds to the BEC regime and $0.2 < C_F < 0.8$ corresponds to the BEC-BCS crossover regime. 
 
At zero temperature, the coupled mean field BCS-like equations for the gap parameter and the chemical potential adopted as a reference model in the present work are able to describe in a qualitative way the BCS-BEC crossover when the density is reduced at fixed interlayer
 distance\cite{PhysRevB.89.060502,Zareniasuperfluidity:2014,PhysRevLett.110.146803}. At very low densities and at smaller enough interlayer distance
 the electron-hole system is in the very strong-coupling regime: the gap is large in units of the non interacting Fermi energy, the condensate fraction is close to one, the chemical potential becomes strongly negative and it approaches minus half of the binging energy of the two-body  electron-hole problem, which is the bound state energy of an isolated exciton. Moreover, in
 the BEC limit the radius of the electron-hole Cooper pairs approach the radius of the exciton in the isolated exciton-limit (see also Ref. [\onlinecite{Pablo2017}] for QMC simulations in the BEC regime).  Therefore, the extreme BEC limit realized at very low densities  in our electron-hole system at T=0 is in correspondence with the pure, weakly interacting, excitonic system analyzed in detail in Ref. [\onlinecite{PhysRevB.96.014505}]. On the other hand, a direct quantitative comparison between the results of our work and the results of Ref. [\onlinecite{PhysRevB.96.014505}] is not possible and beyond the scope of the present paper, being different the model interaction between carriers and the way to include the Coulomb screening in the two approaches.
 
 To highlight the anisotropy of the gap function, we provide three contour plots of $\Delta({\boldsymbol{k}})$ in Fig. \ref{Fig:3} for $d=2$ nm and for the densities marked in Fig. \ref{Fig:2} by solid black dots. We see that in Fig. \ref{Fig:3}(a) that at the density $n \approx 1 \times 10^{12} \text{cm}^{-2}$ the peak in $\Delta({\boldsymbol{k}})$ is centered at $k=0$ in the strong pairing regime of BEC. In Fig. \ref{Fig:3}(b) the maximum $\Delta({\boldsymbol{k}})$ is becoming more broad along the zigzag direction but it is still centered around $k=0$ displaying a remaining bosonic character of the Cooper pairing in the BEC regime but the maximum peak goes to the BEC-BCS crossover regime of weaker coupling along the armchair direction. Fig. \ref{Fig:3}(c) shows the contour plot of $\Delta({\boldsymbol{k}})$ at $n \approx 3.5 \times 10^{12} \text{cm}^{-2}$ where the superfluid gap function along the zigzag direction has its maximum.   
 As one can see in Fig. \ref{Fig:3}(c) our results for $\Delta({\boldsymbol{k}})$ confirm that we are in the BEC-BCS crossover regime along the zigzag direction, while the superfluid gap vanishes along the armchair direction.
 
 \begin{figure*}[ht]
 	\includegraphics[width=18.3cm]{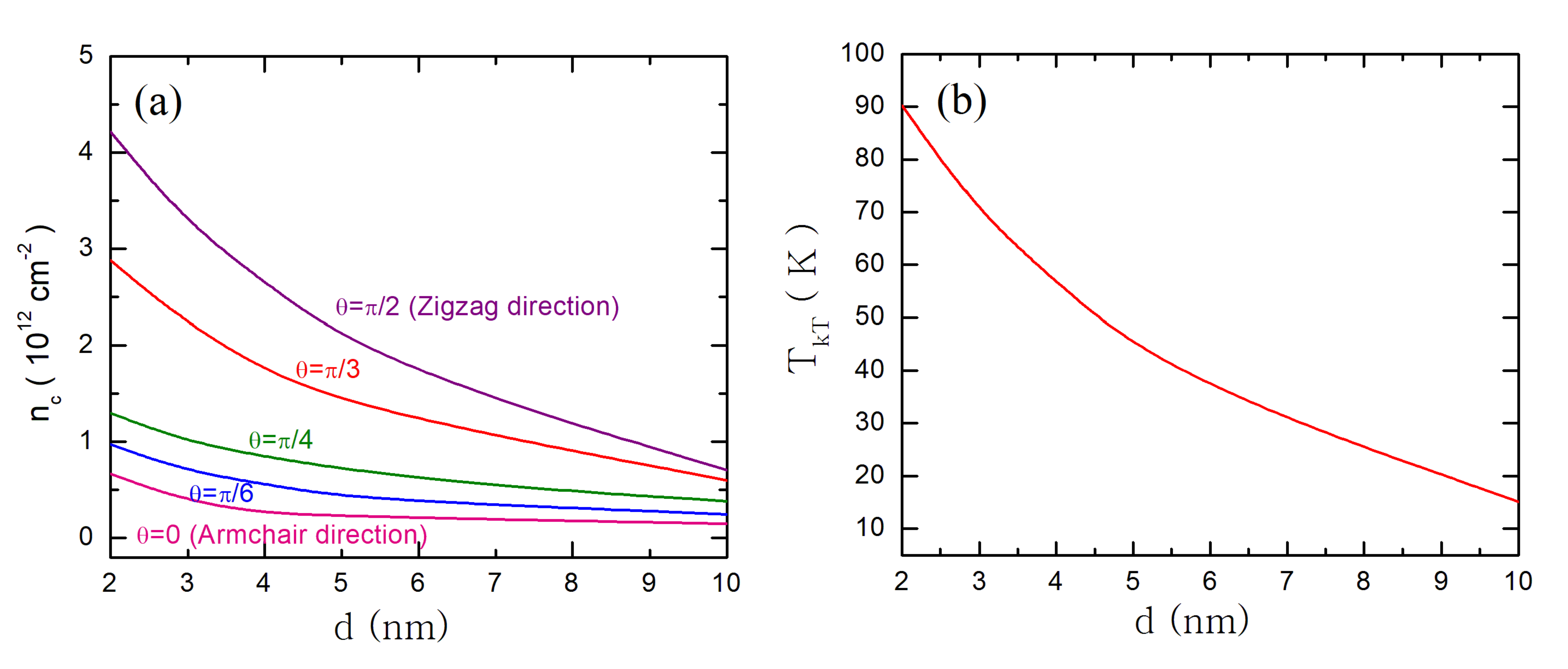}
 	\caption{(a) Maximum density n$_c$ for several directions as a function of the h-BN effective barrier thickness $d$. (b) Maximum T$_{KT}$ as a function of the h-BN effective barrier thickness $d$.}
 	\label{Fig:4}
 \end{figure*}
 
 In Fig. \ref{Fig:4}(a), we show the onset density $n_c$, \textit{i.e.} the density at which the superfluidity is killed, as a function of $d$ for different directions of two alignment phosphorene layers. At a fixed $d$, the onset density is larger along the zigzag ($\theta=\pi/2$) direction. Its value in this direction is about five times the one for the armchair ($\theta=0$) direction as $d$ decreases. We note that once a superfluid condensate is established in one direction, it forms a 2D coherent state and not only in a specific edge.  
The superfluid density $ \rho_s(T_{KT})$ becomes a tensor in the case of an anisotropic system \cite{PhysRevB.96.014505}. The elements of the superfluid density tensor at BCS level for an anisotropic energy band are given by \cite{PhysRevB.78.140502}
	
	\begin{equation}
	\begin{aligned}
	\rho^{x/y}_s(0)&= \int \frac{d^2 \boldsymbol{k}}{(2 \pi)^2} \frac{\partial^2 \xi (\boldsymbol{k})}{\partial k^2_{x/y}} \bigg[1-\frac{\xi(\boldsymbol{k})-\mu}{E(\boldsymbol{k})}\text{tanh}\big(\frac{E(\boldsymbol{k})}{2T}\big)\bigg]\\
	&+2\int \frac{d^2 \boldsymbol{k}}{(2 \pi)^2} \bigg(\frac{\partial \xi(\boldsymbol{k})}{\partial k_{x/y}}\bigg)^2 \frac{\partial f (E(\boldsymbol{k}))}{\partial E(\boldsymbol{k})},
		\end{aligned}
	\end{equation}
	
	\noindent where $f(x)=[1+\text{exp}(x/T)]^{-1}$ is the Fermi Dirac distribution function. At $T=0$ the superfluid density tensor can be simplified to \cite{PhysRevB.78.140502}
	
	\begin{equation}
		\rho^{x/y}_s(0)= 2 \int \frac{d^2 \boldsymbol{k}}{(2 \pi)^2} \frac{ \nu^2({\boldsymbol{k}})}{m_{x/y}}= \frac{n}{m_{x/y}}.
		\label{eq:18}
	\end{equation}
	
	Using the transformations (\ref{eq:trans}) one can write $\rho_s(0)=n R(\theta)$. We determine an upper bound to the superfluid transition temperature using the Kosterlitz-Thouless relation

	\begin{equation}
	T_{KT}=\frac{\pi}{4} \rho_s(T_{KT}),
	\label{eq:17}
	\end{equation}
	
	 The maximum $T_{KT}$ at the onset densities is plotted in Fig. \ref{Fig:4}(b) as a function of $d$. This is the highest temperature, below which we find superfluidity. Notice that these maximum values for the KT transition temperatures are larger than those predicted for double few-layer sheets of graphene\cite{Zareniasuperfluidity:2014}. In contrast with the results in Ref. [\onlinecite{PhysRevB.96.014505}], we find the critical temperature and the onset superfluid density decrease when the interlayer separation $d$ increases. This results from the factor $\exp(-qd)$ in the Fourier transformed interlayer Coulomb interaction, see Eq. (\ref{eq:epsilon}), which ensures suppression of superfluity in the weakly interacting regime when $qd \gg 1$.
	
	\section{conclusion} \label{conclude}
	In summary, we studied the occurrence of excitonic superfluidity  in parallel e-h phosphorene monolayers. We first generalized the mean-field and RPA approach for the gap function to strongly anisotropic double-layer systems such as phosphorene at zero temperature. We showed that the gap function is anisotropic, and depends on the direction of $\boldsymbol{k}$. The origin of this anisotropy can be traced back to the anisotropic effective mass in the free-electron model for phosphorene monolayers. We found that due to the larger effective mass in the zigzag direction the superfluid gap is about twice larger along this direction than along the armchair direction. We estimated the KT transition temperature with maximum value up to $\sim 90$ K in double-layer phosphorene with onset carrier densities as high as $4 \times 10^{12}$ cm$^{-2}$.  These values suggest double electron-hole phosphorene monolayers separated by  \BN\ layers as an experimentally tunable system to observe anisotropic electron-hole superfluid at high temperature and carrier densities, with transition temperatures that are higher than the ones predicted for double graphene bilayers.  The combination of extremely thin barriers, large effective masses and strong pairing attraction make the double phosphorene monolayers ideal for observing high temperature superfluidity.   the separation $d$ between layers increases,The anisotropy in the electron-hole superfluidity can be detected in Coulomb drag experiments by rotating one of the phosphorene sheets with respect to the other. 
\\	
	\section*{Acknowledgements}
	We thank David Neilson for helpful discussions. This work was partially supported by the Flemish Science Foundation (FWO-Vl)  and the Methusalem program of the Flemish government and Iran Ministry of Science, Research and Technology.


\begin{thebibliography}{56}%
	\makeatletter
	\providecommand \@ifxundefined [1]{%
		\@ifx{#1\undefined}
	}%
	\providecommand \@ifnum [1]{%
		\ifnum #1\expandafter \@firstoftwo
		\else \expandafter \@secondoftwo
		\fi
	}%
	\providecommand \@ifx [1]{%
		\ifx #1\expandafter \@firstoftwo
		\else \expandafter \@secondoftwo
		\fi
	}%
	\providecommand \natexlab [1]{#1}%
	\providecommand \enquote  [1]{``#1''}%
	\providecommand \bibnamefont  [1]{#1}%
	\providecommand \bibfnamefont [1]{#1}%
	\providecommand \citenamefont [1]{#1}%
	\providecommand \href@noop [0]{\@secondoftwo}%
	\providecommand \href [0]{\begingroup \@sanitize@url \@href}%
	\providecommand \@href[1]{\@@startlink{#1}\@@href}%
	\providecommand \@@href[1]{\endgroup#1\@@endlink}%
	\providecommand \@sanitize@url [0]{\catcode `\\12\catcode `\$12\catcode
		`\&12\catcode `\#12\catcode `\^12\catcode `\_12\catcode `\%12\relax}%
	\providecommand \@@startlink[1]{}%
	\providecommand \@@endlink[0]{}%
	\providecommand \url  [0]{\begingroup\@sanitize@url \@url }%
	\providecommand \@url [1]{\endgroup\@href {#1}{\urlprefix }}%
	\providecommand \urlprefix  [0]{URL }%
	\providecommand \Eprint [0]{\href }%
	\providecommand \doibase [0]{http://dx.doi.org/}%
	\providecommand \selectlanguage [0]{\@gobble}%
	\providecommand \bibinfo  [0]{\@secondoftwo}%
	\providecommand \bibfield  [0]{\@secondoftwo}%
	\providecommand \translation [1]{[#1]}%
	\providecommand \BibitemOpen [0]{}%
	\providecommand \bibitemStop [0]{}%
	\providecommand \bibitemNoStop [0]{.\EOS\space}%
	\providecommand \EOS [0]{\spacefactor3000\relax}%
	\providecommand \BibitemShut  [1]{\csname bibitem#1\endcsname}%
	\let\auto@bib@innerbib\@empty
	\bibitem [{\citenamefont {Keyes}(1953)}]{PhysRev.92.580}%
	\BibitemOpen
	\bibfield  {author} {\bibinfo {author} {\bibfnamefont {R.~W.}\ \bibnamefont
			{Keyes}},\ }\href {\doibase 10.1103/PhysRev.92.580} {\bibfield  {journal}
		{\bibinfo  {journal} {Phys. Rev.}\ }\textbf {\bibinfo {volume} {92}},\
		\bibinfo {pages} {580} (\bibinfo {year} {1953})}\BibitemShut {NoStop}%
	\bibitem [{\citenamefont {Jamieson}(1963)}]{John:sc63}%
	\BibitemOpen
	\bibfield  {author} {\bibinfo {author} {\bibfnamefont {J.~C.}\ \bibnamefont
			{Jamieson}},\ }\href {\doibase 10.1126/science.139.3561.1291} {\bibfield
		{journal} {\bibinfo  {journal} {Science}\ }\textbf {\bibinfo {volume}
			{139}},\ \bibinfo {pages} {1291} (\bibinfo {year} {1963})}\BibitemShut
	{NoStop}%
	\bibitem [{\citenamefont {Appalakondaiah}\ \emph {et~al.}(2012)\citenamefont
		{Appalakondaiah}, \citenamefont {Vaitheeswaran}, \citenamefont {Leb\`egue},
		\citenamefont {Christensen},\ and\ \citenamefont
		{Svane}}]{PhysRevB.86.035105}%
	\BibitemOpen
	\bibfield  {author} {\bibinfo {author} {\bibfnamefont {S.}~\bibnamefont
			{Appalakondaiah}}, \bibinfo {author} {\bibfnamefont {G.}~\bibnamefont
			{Vaitheeswaran}}, \bibinfo {author} {\bibfnamefont {S.}~\bibnamefont
			{Leb\`egue}}, \bibinfo {author} {\bibfnamefont {N.~E.}\ \bibnamefont
			{Christensen}}, \ and\ \bibinfo {author} {\bibfnamefont {A.}~\bibnamefont
			{Svane}},\ }\href {\doibase 10.1103/PhysRevB.86.035105} {\bibfield  {journal}
		{\bibinfo  {journal} {Phys. Rev. B}\ }\textbf {\bibinfo {volume} {86}},\
		\bibinfo {pages} {035105} (\bibinfo {year} {2012})}\BibitemShut {NoStop}%
	\bibitem [{\citenamefont {Li}\ and\ \citenamefont {et~al}(2017)}]{Likai:nat17}%
	\BibitemOpen
	\bibfield  {author} {\bibinfo {author} {\bibfnamefont {L.}~\bibnamefont
			{Li}}\ and\ \bibinfo {author} {\bibnamefont {et~al}},\ }\href {\doibase
		10.1038/nnano.2016.171} {\bibfield  {journal} {\bibinfo  {journal} {Nat.
				Nanotechnol.}\ }\textbf {\bibinfo {volume} {12}},\ \bibinfo {pages} {21}
		(\bibinfo {year} {2017})}\BibitemShut {NoStop}%
	\bibitem [{\citenamefont {Xia}\ \emph {et~al.}(2014)\citenamefont {Xia},
		\citenamefont {Wang},\ and\ \citenamefont {Jia}}]{Xia:nat13}%
	\BibitemOpen
	\bibfield  {author} {\bibinfo {author} {\bibfnamefont {F.}~\bibnamefont
			{Xia}}, \bibinfo {author} {\bibfnamefont {H.}~\bibnamefont {Wang}}, \ and\
		\bibinfo {author} {\bibfnamefont {Y.}~\bibnamefont {Jia}},\ }\href {\doibase
		10.1038/ncomms5458} {\bibfield  {journal} {\bibinfo  {journal} {Nat.
				Commun.}\ }\textbf {\bibinfo {volume} {5}},\ \bibinfo {pages} {4458}
		(\bibinfo {year} {2014})}\BibitemShut {NoStop}%
	\bibitem [{\citenamefont {Carvalho}\ \emph {et~al.}(2016)\citenamefont
		{Carvalho}, \citenamefont {Wang}, \citenamefont {Zhu}, \citenamefont {Rodin},
		\citenamefont {Su},\ and\ \citenamefont {Neto}}]{Neto:rev2016}%
	\BibitemOpen
	\bibfield  {author} {\bibinfo {author} {\bibfnamefont {N.}~\bibnamefont
			{Carvalho}}, \bibinfo {author} {\bibfnamefont {M.}~\bibnamefont {Wang}},
		\bibinfo {author} {\bibfnamefont {X.}~\bibnamefont {Zhu}}, \bibinfo {author}
		{\bibfnamefont {A.~S.}\ \bibnamefont {Rodin}}, \bibinfo {author}
		{\bibfnamefont {H.}~\bibnamefont {Su}}, \ and\ \bibinfo {author}
		{\bibfnamefont {A.~H.~C.}\ \bibnamefont {Neto}},\ }\href
	{https://www.nature.com/articles/natrevmats201661} {\bibfield  {journal}
		{\bibinfo  {journal} {Nature Reviews Materials}\ }\textbf {\bibinfo {volume}
			{1}} (\bibinfo {year} {2016})}\BibitemShut {NoStop}%
	\bibitem [{\citenamefont {Tran}\ \emph {et~al.}(2014)\citenamefont {Tran},
		\citenamefont {Soklaski}, \citenamefont {Liang},\ and\ \citenamefont
		{Yang}}]{PhysRevB.89.235319}%
	\BibitemOpen
	\bibfield  {author} {\bibinfo {author} {\bibfnamefont {V.}~\bibnamefont
			{Tran}}, \bibinfo {author} {\bibfnamefont {R.}~\bibnamefont {Soklaski}},
		\bibinfo {author} {\bibfnamefont {Y.}~\bibnamefont {Liang}}, \ and\ \bibinfo
		{author} {\bibfnamefont {L.}~\bibnamefont {Yang}},\ }\href {\doibase
		10.1103/PhysRevB.89.235319} {\bibfield  {journal} {\bibinfo  {journal} {Phys.
				Rev. B}\ }\textbf {\bibinfo {volume} {89}},\ \bibinfo {pages} {235319}
		(\bibinfo {year} {2014})}\BibitemShut {NoStop}%
	\bibitem [{\citenamefont {Liu}\ \emph {et~al.}(2014)\citenamefont {Liu},
		\citenamefont {Neal}, \citenamefont {Zhu}, \citenamefont {Luo}, \citenamefont
		{Xu}, \citenamefont {Tom{\'a}nek},\ and\ \citenamefont {Ye}}]{liu:acs14}%
	\BibitemOpen
	\bibfield  {author} {\bibinfo {author} {\bibfnamefont {H.}~\bibnamefont
			{Liu}}, \bibinfo {author} {\bibfnamefont {A.~T.}\ \bibnamefont {Neal}},
		\bibinfo {author} {\bibfnamefont {Z.}~\bibnamefont {Zhu}}, \bibinfo {author}
		{\bibfnamefont {Z.}~\bibnamefont {Luo}}, \bibinfo {author} {\bibfnamefont
			{X.}~\bibnamefont {Xu}}, \bibinfo {author} {\bibfnamefont {D.}~\bibnamefont
			{Tom{\'a}nek}}, \ and\ \bibinfo {author} {\bibfnamefont {P.~D.}\ \bibnamefont
			{Ye}},\ }\href {\doibase 10.1021/nn501226z} {\bibfield  {journal} {\bibinfo
			{journal} {ACS nano}\ }\textbf {\bibinfo {volume} {8}},\ \bibinfo {pages}
		{4033} (\bibinfo {year} {2014})}\BibitemShut {NoStop}%
	\bibitem [{\citenamefont {Ostahie}\ and\ \citenamefont
		{Aldea}(2016)}]{PhysRevB.93.075408}%
	\BibitemOpen
	\bibfield  {author} {\bibinfo {author} {\bibfnamefont {B.}~\bibnamefont
			{Ostahie}}\ and\ \bibinfo {author} {\bibfnamefont {A.}~\bibnamefont
			{Aldea}},\ }\href {\doibase 10.1103/PhysRevB.93.075408} {\bibfield  {journal}
		{\bibinfo  {journal} {Phys. Rev. B}\ }\textbf {\bibinfo {volume} {93}},\
		\bibinfo {pages} {075408} (\bibinfo {year} {2016})}\BibitemShut {NoStop}%
	\bibitem [{\citenamefont {Padilha}\ \emph {et~al.}(2015)\citenamefont
		{Padilha}, \citenamefont {Fazzio},\ and\ \citenamefont
		{da~Silva}}]{PhysRevLett.114.066803}%
	\BibitemOpen
	\bibfield  {author} {\bibinfo {author} {\bibfnamefont {J.~E.}\ \bibnamefont
			{Padilha}}, \bibinfo {author} {\bibfnamefont {A.}~\bibnamefont {Fazzio}}, \
		and\ \bibinfo {author} {\bibfnamefont {A.~J.~R.}\ \bibnamefont {da~Silva}},\
	}\href {\doibase 10.1103/PhysRevLett.114.066803} {\bibfield  {journal}
	{\bibinfo  {journal} {Phys. Rev. Lett.}\ }\textbf {\bibinfo {volume} {114}},\
	\bibinfo {pages} {066803} (\bibinfo {year} {2015})}\BibitemShut {NoStop}%
\bibitem [{\citenamefont {\ifmmode \mbox{\c{C}}\else \c{C}\fi{}ak\ifmmode
		\imath~\else \i\fi{}r}\ \emph {et~al.}(2014)\citenamefont {\ifmmode
		\mbox{\c{C}}\else \c{C}\fi{}ak\ifmmode \imath~\else \i\fi{}r}, \citenamefont
	{Sahin},\ and\ \citenamefont {Peeters}}]{PhysRevB.90.205421}%
\BibitemOpen
\bibfield  {author} {\bibinfo {author} {\bibfnamefont {D.}~\bibnamefont
		{\ifmmode \mbox{\c{C}}\else \c{C}\fi{}ak\ifmmode \imath~\else \i\fi{}r}},
	\bibinfo {author} {\bibfnamefont {H.}~\bibnamefont {Sahin}}, \ and\ \bibinfo
	{author} {\bibfnamefont {F.~M.}\ \bibnamefont {Peeters}},\ }\href {\doibase
	10.1103/PhysRevB.90.205421} {\bibfield  {journal} {\bibinfo  {journal} {Phys.
			Rev. B}\ }\textbf {\bibinfo {volume} {90}},\ \bibinfo {pages} {205421}
	(\bibinfo {year} {2014})}\BibitemShut {NoStop}%
\bibitem [{\citenamefont {Taghizadeh~Sisakht}\ \emph
	{et~al.}(2016)\citenamefont {Taghizadeh~Sisakht}, \citenamefont {Fazileh},
	\citenamefont {Zare}, \citenamefont {Zarenia},\ and\ \citenamefont
	{Peeters}}]{PhysRevB.94.085417}%
\BibitemOpen
\bibfield  {author} {\bibinfo {author} {\bibfnamefont {E.}~\bibnamefont
		{Taghizadeh~Sisakht}}, \bibinfo {author} {\bibfnamefont {F.}~\bibnamefont
		{Fazileh}}, \bibinfo {author} {\bibfnamefont {M.~H.}\ \bibnamefont {Zare}},
	\bibinfo {author} {\bibfnamefont {M.}~\bibnamefont {Zarenia}}, \ and\
	\bibinfo {author} {\bibfnamefont {F.~M.}\ \bibnamefont {Peeters}},\ }\href
{\doibase 10.1103/PhysRevB.94.085417} {\bibfield  {journal} {\bibinfo
		{journal} {Phys. Rev. B}\ }\textbf {\bibinfo {volume} {94}},\ \bibinfo
	{pages} {085417} (\bibinfo {year} {2016})}\BibitemShut {NoStop}%
\bibitem [{\citenamefont {Wang}\ \emph {et~al.}(2015)\citenamefont {Wang},
	\citenamefont {Jones}, \citenamefont {Seyler}, \citenamefont {Tran},
	\citenamefont {Jia}, \citenamefont {Zhao}, \citenamefont {Wang},
	\citenamefont {Yang}, \citenamefont {Xu},\ and\ \citenamefont
	{Xia}}]{Xiaomu:nat15}%
\BibitemOpen
\bibfield  {author} {\bibinfo {author} {\bibfnamefont {X.}~\bibnamefont
		{Wang}}, \bibinfo {author} {\bibfnamefont {A.~M.}\ \bibnamefont {Jones}},
	\bibinfo {author} {\bibfnamefont {K.~L.}\ \bibnamefont {Seyler}}, \bibinfo
	{author} {\bibfnamefont {V.}~\bibnamefont {Tran}}, \bibinfo {author}
	{\bibfnamefont {Y.}~\bibnamefont {Jia}}, \bibinfo {author} {\bibfnamefont
		{H.}~\bibnamefont {Zhao}}, \bibinfo {author} {\bibfnamefont {H.}~\bibnamefont
		{Wang}}, \bibinfo {author} {\bibfnamefont {L.}~\bibnamefont {Yang}}, \bibinfo
	{author} {\bibfnamefont {X.}~\bibnamefont {Xu}}, \ and\ \bibinfo {author}
	{\bibfnamefont {F.}~\bibnamefont {Xia}},\ }\href {\doibase
	10.1038/nnano.2015.71} {\bibfield  {journal} {\bibinfo  {journal} {Nat.
			Nanotechnol.}\ }\textbf {\bibinfo {volume} {10}},\ \bibinfo {pages} {517}
	(\bibinfo {year} {2015})}\BibitemShut {NoStop}%
\bibitem [{\citenamefont {Saberi-Pouya}\ \emph
	{et~al.}(2017{\natexlab{a}})\citenamefont {Saberi-Pouya}, \citenamefont
	{Vazifehshenas}, \citenamefont {Salavati-fard}, \citenamefont {Farmanbar},\
	and\ \citenamefont {Peeters}}]{Samira:Conductivity2017}%
\BibitemOpen
\bibfield  {author} {\bibinfo {author} {\bibfnamefont {S.}~\bibnamefont
		{Saberi-Pouya}}, \bibinfo {author} {\bibfnamefont {T.}~\bibnamefont
		{Vazifehshenas}}, \bibinfo {author} {\bibfnamefont {T.}~\bibnamefont
		{Salavati-fard}}, \bibinfo {author} {\bibfnamefont {M.}~\bibnamefont
		{Farmanbar}}, \ and\ \bibinfo {author} {\bibfnamefont {F.~M.}\ \bibnamefont
		{Peeters}},\ }\href {\doibase 10.1103/PhysRevB.96.075411} {\bibfield
	{journal} {\bibinfo  {journal} {Phys. Rev. B}\ }\textbf {\bibinfo {volume}
		{96}},\ \bibinfo {pages} {075411} (\bibinfo {year}
	{2017}{\natexlab{a}})}\BibitemShut {NoStop}%
\bibitem [{\citenamefont {Qiao}\ \emph {et~al.}(2014)\citenamefont {Qiao},
	\citenamefont {Kong}, \citenamefont {Hu}, \citenamefont {Yang},\ and\
	\citenamefont {Ji}}]{qiao2014high}%
\BibitemOpen
\bibfield  {author} {\bibinfo {author} {\bibfnamefont {J.}~\bibnamefont
		{Qiao}}, \bibinfo {author} {\bibfnamefont {X.}~\bibnamefont {Kong}}, \bibinfo
	{author} {\bibfnamefont {Z.-X.}\ \bibnamefont {Hu}}, \bibinfo {author}
	{\bibfnamefont {F.}~\bibnamefont {Yang}}, \ and\ \bibinfo {author}
	{\bibfnamefont {W.}~\bibnamefont {Ji}},\ }\href {\doibase
	doi:10.1038/ncomms5458} {\bibfield  {journal} {\bibinfo  {journal} {Nat.
			Commun.}\ }\textbf {\bibinfo {volume} {5}},\ \bibinfo {pages} {4458 }
	(\bibinfo {year} {2014})}\BibitemShut {NoStop}%
\bibitem [{\citenamefont {Prishchenko}\ \emph {et~al.}(2017)\citenamefont
	{Prishchenko}, \citenamefont {Mazurenko}, \citenamefont {Katsnelson},\ and\
	\citenamefont {Rudenko}}]{2053-1583-4-2-025064}%
\BibitemOpen
\bibfield  {author} {\bibinfo {author} {\bibfnamefont {D.~A.}\ \bibnamefont
		{Prishchenko}}, \bibinfo {author} {\bibfnamefont {V.~G.}\ \bibnamefont
		{Mazurenko}}, \bibinfo {author} {\bibfnamefont {M.~I.}\ \bibnamefont
		{Katsnelson}}, \ and\ \bibinfo {author} {\bibfnamefont {A.~N.}\ \bibnamefont
		{Rudenko}},\ }\href {http://stacks.iop.org/2053-1583/4/i=2/a=025064}
{\bibfield  {journal} {\bibinfo  {journal} {2D Materials}\ }\textbf {\bibinfo
		{volume} {4}},\ \bibinfo {pages} {025064} (\bibinfo {year}
	{2017})}\BibitemShut {NoStop}%
\bibitem [{\citenamefont {Saberi-Pouya}\ \emph
	{et~al.}(2017{\natexlab{b}})\citenamefont {Saberi-Pouya}, \citenamefont
	{Vazifehshenas}, \citenamefont {Salavati-fard},\ and\ \citenamefont
	{Farmanbar}}]{Samira:SOphonon2017}%
\BibitemOpen
\bibfield  {author} {\bibinfo {author} {\bibfnamefont {S.}~\bibnamefont
		{Saberi-Pouya}}, \bibinfo {author} {\bibfnamefont {T.}~\bibnamefont
		{Vazifehshenas}}, \bibinfo {author} {\bibfnamefont {T.}~\bibnamefont
		{Salavati-fard}}, \ and\ \bibinfo {author} {\bibfnamefont {M.}~\bibnamefont
		{Farmanbar}},\ }\href {\doibase 10.1103/PhysRevB.96.115402} {\bibfield
	{journal} {\bibinfo  {journal} {Phys. Rev. B}\ }\textbf {\bibinfo {volume}
		{96}},\ \bibinfo {pages} {115402} (\bibinfo {year}
	{2017}{\natexlab{b}})}\BibitemShut {NoStop}%
\bibitem [{\citenamefont {Low}\ \emph {et~al.}(2014{\natexlab{a}})\citenamefont
	{Low}, \citenamefont {Rold\'an}, \citenamefont {Wang}, \citenamefont {Xia},
	\citenamefont {Avouris}, \citenamefont {Moreno},\ and\ \citenamefont
	{Guinea}}]{Low:prl14}%
\BibitemOpen
\bibfield  {author} {\bibinfo {author} {\bibfnamefont {T.}~\bibnamefont
		{Low}}, \bibinfo {author} {\bibfnamefont {R.}~\bibnamefont {Rold\'an}},
	\bibinfo {author} {\bibfnamefont {H.}~\bibnamefont {Wang}}, \bibinfo {author}
	{\bibfnamefont {F.}~\bibnamefont {Xia}}, \bibinfo {author} {\bibfnamefont
		{P.}~\bibnamefont {Avouris}}, \bibinfo {author} {\bibfnamefont {L.~M.}\
		\bibnamefont {Moreno}}, \ and\ \bibinfo {author} {\bibfnamefont
		{F.}~\bibnamefont {Guinea}},\ }\href {\doibase
	10.1103/PhysRevLett.113.106802} {\bibfield  {journal} {\bibinfo  {journal}
		{Phys. Rev. Lett.}\ }\textbf {\bibinfo {volume} {113}},\ \bibinfo {pages}
	{106802} (\bibinfo {year} {2014}{\natexlab{a}})}\BibitemShut {NoStop}%
\bibitem [{\citenamefont {Jin}\ \emph {et~al.}(2015)\citenamefont {Jin},
	\citenamefont {Rold\'an}, \citenamefont {Katsnelson},\ and\ \citenamefont
	{Yuan}}]{Jin:prb15}%
\BibitemOpen
\bibfield  {author} {\bibinfo {author} {\bibfnamefont {F.}~\bibnamefont
		{Jin}}, \bibinfo {author} {\bibfnamefont {R.}~\bibnamefont {Rold\'an}},
	\bibinfo {author} {\bibfnamefont {M.~I.}\ \bibnamefont {Katsnelson}}, \ and\
	\bibinfo {author} {\bibfnamefont {S.}~\bibnamefont {Yuan}},\ }\href {\doibase
	10.1103/PhysRevB.92.115440} {\bibfield  {journal} {\bibinfo  {journal} {Phys.
			Rev. B}\ }\textbf {\bibinfo {volume} {92}},\ \bibinfo {pages} {115440}
	(\bibinfo {year} {2015})}\BibitemShut {NoStop}%
\bibitem [{\citenamefont {Rodin}\ and\ \citenamefont
	{Castro~Neto}(2015)}]{Rodin:prb15}%
\BibitemOpen
\bibfield  {author} {\bibinfo {author} {\bibfnamefont {A.~S.}\ \bibnamefont
		{Rodin}}\ and\ \bibinfo {author} {\bibfnamefont {A.~H.}\ \bibnamefont
		{Castro~Neto}},\ }\href {\doibase 10.1103/PhysRevB.91.075422} {\bibfield
	{journal} {\bibinfo  {journal} {Phys. Rev. B}\ }\textbf {\bibinfo {volume}
		{91}},\ \bibinfo {pages} {075422} (\bibinfo {year} {2015})}\BibitemShut
{NoStop}%
\bibitem [{\citenamefont {Saberi-Pouya}\ \emph {et~al.}(2016)\citenamefont
	{Saberi-Pouya}, \citenamefont {Vazifehshenas}, \citenamefont {Farmanbar},\
	and\ \citenamefont {Salavati-fard}}]{Samira:drag2016}%
\BibitemOpen
\bibfield  {author} {\bibinfo {author} {\bibfnamefont {S.}~\bibnamefont
		{Saberi-Pouya}}, \bibinfo {author} {\bibfnamefont {T.}~\bibnamefont
		{Vazifehshenas}}, \bibinfo {author} {\bibfnamefont {M.}~\bibnamefont
		{Farmanbar}}, \ and\ \bibinfo {author} {\bibfnamefont {T.}~\bibnamefont
		{Salavati-fard}},\ }\href {http://stacks.iop.org/0953-8984/28/i=28/a=285301}
{\bibfield  {journal} {\bibinfo  {journal} {Journal of Physics: Condensed
			Matter}\ }\textbf {\bibinfo {volume} {28}},\ \bibinfo {pages} {285301}
	(\bibinfo {year} {2016})}\BibitemShut {NoStop}%
\bibitem [{\citenamefont {Island}\ \emph {et~al.}(2015)\citenamefont {Island},
	\citenamefont {Steele}, \citenamefont {van~der Zant},\ and\ \citenamefont
	{Castellanos-Gomez}}]{2053-1583-2-1-011002}%
\BibitemOpen
\bibfield  {author} {\bibinfo {author} {\bibfnamefont {J.~O.}\ \bibnamefont
		{Island}}, \bibinfo {author} {\bibfnamefont {G.~A.}\ \bibnamefont {Steele}},
	\bibinfo {author} {\bibfnamefont {H.~S.~J.}\ \bibnamefont {van~der Zant}}, \
	and\ \bibinfo {author} {\bibfnamefont {A.}~\bibnamefont
		{Castellanos-Gomez}},\ }\href
{http://stacks.iop.org/2053-1583/2/i=1/a=011002} {\bibfield  {journal}
	{\bibinfo  {journal} {2D Materials}\ }\textbf {\bibinfo {volume} {2}},\
	\bibinfo {pages} {011002} (\bibinfo {year} {2015})}\BibitemShut {NoStop}%
\bibitem [{\citenamefont {Favron}\ and\ \citenamefont
	{et~al}(2015)}]{Favron:nat15}%
\BibitemOpen
\bibfield  {author} {\bibinfo {author} {\bibfnamefont {A.}~\bibnamefont
		{Favron}}\ and\ \bibinfo {author} {\bibnamefont {et~al}},\ }\href {\doibase
	10.1038/nmat4299} {\bibfield  {journal} {\bibinfo  {journal} {Nature
			Materials}\ }\textbf {\bibinfo {volume} {14}},\ \bibinfo {pages} {826}
	(\bibinfo {year} {2015})}\BibitemShut {NoStop}%
\bibitem [{\citenamefont {Cao}\ and\ \citenamefont
	{et~al.}(2015)}]{Cao:nanol15}%
\BibitemOpen
\bibfield  {author} {\bibinfo {author} {\bibfnamefont {Y.}~\bibnamefont
		{Cao}}\ and\ \bibinfo {author} {\bibnamefont {et~al.}},\ }\href {\doibase
	10.1021/acs.nanolett.5b00648} {\bibfield  {journal} {\bibinfo  {journal}
		{Nano Lett.}\ }\textbf {\bibinfo {volume} {15}},\ \bibinfo {pages} {4914}
	(\bibinfo {year} {2015})}\BibitemShut {NoStop}%
\bibitem [{\citenamefont {Gorbachev}\ \emph {et~al.}(2012)\citenamefont
	{Gorbachev}, \citenamefont {Geim}, \citenamefont {Katsnelson}, \citenamefont
	{Novoselov}, \citenamefont {Tudorovskiy}, \citenamefont {Grigorieva},
	\citenamefont {MacDonald}, \citenamefont {Morozov}, \citenamefont {Watanabe},
	\citenamefont {Taniguchi},\ and\ \citenamefont
	{Ponomarenko}}]{Gorbachev:natphys12}%
\BibitemOpen
\bibfield  {author} {\bibinfo {author} {\bibfnamefont {R.~V.}\ \bibnamefont
		{Gorbachev}}, \bibinfo {author} {\bibfnamefont {A.~K.}\ \bibnamefont {Geim}},
	\bibinfo {author} {\bibfnamefont {M.~I.}\ \bibnamefont {Katsnelson}},
	\bibinfo {author} {\bibfnamefont {K.~S.}\ \bibnamefont {Novoselov}}, \bibinfo
	{author} {\bibfnamefont {T.}~\bibnamefont {Tudorovskiy}}, \bibinfo {author}
	{\bibfnamefont {I.~V.}\ \bibnamefont {Grigorieva}}, \bibinfo {author}
	{\bibfnamefont {A.~H.}\ \bibnamefont {MacDonald}}, \bibinfo {author}
	{\bibfnamefont {S.~V.}\ \bibnamefont {Morozov}}, \bibinfo {author}
	{\bibfnamefont {K.}~\bibnamefont {Watanabe}}, \bibinfo {author}
	{\bibfnamefont {T.}~\bibnamefont {Taniguchi}}, \ and\ \bibinfo {author}
	{\bibfnamefont {L.~A.}\ \bibnamefont {Ponomarenko}},\ }\href {\doibase
	10.1038/nphys2441} {\bibfield  {journal} {\bibinfo  {journal} {Nature
			Physics}\ }\textbf {\bibinfo {volume} {8}},\ \bibinfo {pages} {896} (\bibinfo
	{year} {2012})}\BibitemShut {NoStop}%
\bibitem [{\citenamefont {Saslow}(1973)}]{PhysRevLett.31.870}%
\BibitemOpen
\bibfield  {author} {\bibinfo {author} {\bibfnamefont {W.~M.}\ \bibnamefont
		{Saslow}},\ }\href {\doibase 10.1103/PhysRevLett.31.870} {\bibfield
	{journal} {\bibinfo  {journal} {Phys. Rev. Lett.}\ }\textbf {\bibinfo
		{volume} {31}},\ \bibinfo {pages} {870} (\bibinfo {year} {1973})}\BibitemShut
{NoStop}%
\bibitem [{\citenamefont {Conti}\ \emph {et~al.}(1998)\citenamefont {Conti},
	\citenamefont {Vignale},\ and\ \citenamefont
	{MacDonald}}]{PhysRevB.57.R6846}%
\BibitemOpen
\bibfield  {author} {\bibinfo {author} {\bibfnamefont {S.}~\bibnamefont
		{Conti}}, \bibinfo {author} {\bibfnamefont {G.}~\bibnamefont {Vignale}}, \
	and\ \bibinfo {author} {\bibfnamefont {A.~H.}\ \bibnamefont {MacDonald}},\
}\href {\doibase 10.1103/PhysRevB.57.R6846} {\bibfield  {journal} {\bibinfo
	{journal} {Phys. Rev. B}\ }\textbf {\bibinfo {volume} {57}},\ \bibinfo
{pages} {R6846} (\bibinfo {year} {1998})}\BibitemShut {NoStop}%
\bibitem [{\citenamefont {Benfatto}\ \emph {et~al.}(2008)\citenamefont
	{Benfatto}, \citenamefont {Capone}, \citenamefont {Caprara}, \citenamefont
	{Castellani},\ and\ \citenamefont {Di~Castro}}]{PhysRevB.78.140502}%
\BibitemOpen
\bibfield  {author} {\bibinfo {author} {\bibfnamefont {L.}~\bibnamefont
		{Benfatto}}, \bibinfo {author} {\bibfnamefont {M.}~\bibnamefont {Capone}},
	\bibinfo {author} {\bibfnamefont {S.}~\bibnamefont {Caprara}}, \bibinfo
	{author} {\bibfnamefont {C.}~\bibnamefont {Castellani}}, \ and\ \bibinfo
	{author} {\bibfnamefont {C.}~\bibnamefont {Di~Castro}},\ }\href {\doibase
	10.1103/PhysRevB.78.140502} {\bibfield  {journal} {\bibinfo  {journal} {Phys.
			Rev. B}\ }\textbf {\bibinfo {volume} {78}},\ \bibinfo {pages} {140502}
	(\bibinfo {year} {2008})}\BibitemShut {NoStop}%
\bibitem [{\citenamefont {Neilson}\ \emph {et~al.}(2014)\citenamefont
	{Neilson}, \citenamefont {Perali},\ and\ \citenamefont
	{Hamilton}}]{PhysRevB.89.060502}%
\BibitemOpen
\bibfield  {author} {\bibinfo {author} {\bibfnamefont {D.}~\bibnamefont
		{Neilson}}, \bibinfo {author} {\bibfnamefont {A.}~\bibnamefont {Perali}}, \
	and\ \bibinfo {author} {\bibfnamefont {A.~R.}\ \bibnamefont {Hamilton}},\
}\href {\doibase 10.1103/PhysRevB.89.060502} {\bibfield  {journal} {\bibinfo
	{journal} {Phys. Rev. B}\ }\textbf {\bibinfo {volume} {89}},\ \bibinfo
{pages} {060502} (\bibinfo {year} {2014})}\BibitemShut {NoStop}%
\bibitem [{\citenamefont {Zarenia}\ \emph {et~al.}(2014)\citenamefont
	{Zarenia}, \citenamefont {Perali}, \citenamefont {Neilson},\ and\
	\citenamefont {Peeters}}]{Zareniasuperfluidity:2014}%
\BibitemOpen
\bibfield  {author} {\bibinfo {author} {\bibfnamefont {M.}~\bibnamefont
		{Zarenia}}, \bibinfo {author} {\bibfnamefont {A.}~\bibnamefont {Perali}},
	\bibinfo {author} {\bibfnamefont {D.}~\bibnamefont {Neilson}}, \ and\
	\bibinfo {author} {\bibfnamefont {F.~M.}\ \bibnamefont {Peeters}},\ }\href
{\doibase 10.1038/srep07319} {\bibfield  {journal} {\bibinfo  {journal}
		{Scientific Reports}\ }\textbf {\bibinfo {volume} {4}},\ \bibinfo {pages}
	{7319} (\bibinfo {year} {2014})}\BibitemShut {NoStop}%
\bibitem [{\citenamefont {Croxall}\ \emph {et~al.}(2008)\citenamefont
	{Croxall}, \citenamefont {Das~Gupta}, \citenamefont {Nicoll}, \citenamefont
	{Thangaraj}, \citenamefont {Beere}, \citenamefont {Farrer}, \citenamefont
	{Ritchie},\ and\ \citenamefont {Pepper}}]{PhysRevLett.101.246801}%
\BibitemOpen
\bibfield  {author} {\bibinfo {author} {\bibfnamefont {A.~F.}\ \bibnamefont
		{Croxall}}, \bibinfo {author} {\bibfnamefont {K.}~\bibnamefont {Das~Gupta}},
	\bibinfo {author} {\bibfnamefont {C.~A.}\ \bibnamefont {Nicoll}}, \bibinfo
	{author} {\bibfnamefont {M.}~\bibnamefont {Thangaraj}}, \bibinfo {author}
	{\bibfnamefont {H.~E.}\ \bibnamefont {Beere}}, \bibinfo {author}
	{\bibfnamefont {I.}~\bibnamefont {Farrer}}, \bibinfo {author} {\bibfnamefont
		{D.~A.}\ \bibnamefont {Ritchie}}, \ and\ \bibinfo {author} {\bibfnamefont
		{M.}~\bibnamefont {Pepper}},\ }\href {\doibase
	10.1103/PhysRevLett.101.246801} {\bibfield  {journal} {\bibinfo  {journal}
		{Phys. Rev. Lett.}\ }\textbf {\bibinfo {volume} {101}},\ \bibinfo {pages}
	{246801} (\bibinfo {year} {2008})}\BibitemShut {NoStop}%
\bibitem [{\citenamefont {Lee}\ \emph {et~al.}(2016)\citenamefont {Lee},
	\citenamefont {Xue}, \citenamefont {Dillen}, \citenamefont {Watanabe},
	\citenamefont {Taniguchi},\ and\ \citenamefont
	{Tutuc}}]{PhysRevLett.117.046803}%
\BibitemOpen
\bibfield  {author} {\bibinfo {author} {\bibfnamefont {K.}~\bibnamefont
		{Lee}}, \bibinfo {author} {\bibfnamefont {J.}~\bibnamefont {Xue}}, \bibinfo
	{author} {\bibfnamefont {D.~C.}\ \bibnamefont {Dillen}}, \bibinfo {author}
	{\bibfnamefont {K.}~\bibnamefont {Watanabe}}, \bibinfo {author}
	{\bibfnamefont {T.}~\bibnamefont {Taniguchi}}, \ and\ \bibinfo {author}
	{\bibfnamefont {E.}~\bibnamefont {Tutuc}},\ }\href {\doibase
	10.1103/PhysRevLett.117.046803} {\bibfield  {journal} {\bibinfo  {journal}
		{Phys. Rev. Lett.}\ }\textbf {\bibinfo {volume} {117}},\ \bibinfo {pages}
	{046803} (\bibinfo {year} {2016})}\BibitemShut {NoStop}%
\bibitem [{\citenamefont {Liu}\ \emph {et~al.}(2017)\citenamefont {Liu},
	\citenamefont {Watanabe}, \citenamefont {Taniguchi}, \citenamefont
	{Halperin},\ and\ \citenamefont {Kim}}]{Liu:nat17}%
\BibitemOpen
\bibfield  {author} {\bibinfo {author} {\bibfnamefont {X.}~\bibnamefont
		{Liu}}, \bibinfo {author} {\bibfnamefont {K.}~\bibnamefont {Watanabe}},
	\bibinfo {author} {\bibfnamefont {T.}~\bibnamefont {Taniguchi}}, \bibinfo
	{author} {\bibfnamefont {B.~I.}\ \bibnamefont {Halperin}}, \ and\ \bibinfo
	{author} {\bibfnamefont {P.}~\bibnamefont {Kim}},\ }\href {\doibase
	10.1038/nphys4116} {\bibfield  {journal} {\bibinfo  {journal} {Nature
			Physics}\ }\textbf {\bibinfo {volume} {13}},\ \bibinfo {pages} {746}
	(\bibinfo {year} {2017})}\BibitemShut {NoStop}%
\bibitem [{\citenamefont {Lozovik}\ \emph {et~al.}(2012)\citenamefont
	{Lozovik}, \citenamefont {Ogarkov},\ and\ \citenamefont
	{Sokolik}}]{PhysRevB.86.045429}%
\BibitemOpen
\bibfield  {author} {\bibinfo {author} {\bibfnamefont {Y.~E.}\ \bibnamefont
		{Lozovik}}, \bibinfo {author} {\bibfnamefont {S.~L.}\ \bibnamefont
		{Ogarkov}}, \ and\ \bibinfo {author} {\bibfnamefont {A.~A.}\ \bibnamefont
		{Sokolik}},\ }\href {\doibase 10.1103/PhysRevB.86.045429} {\bibfield
	{journal} {\bibinfo  {journal} {Phys. Rev. B}\ }\textbf {\bibinfo {volume}
		{86}},\ \bibinfo {pages} {045429} (\bibinfo {year} {2012})}\BibitemShut
{NoStop}%
\bibitem [{\citenamefont {Perali}\ \emph {et~al.}(2013)\citenamefont {Perali},
	\citenamefont {Neilson},\ and\ \citenamefont
	{Hamilton}}]{PhysRevLett.110.146803}%
\BibitemOpen
\bibfield  {author} {\bibinfo {author} {\bibfnamefont {A.}~\bibnamefont
		{Perali}}, \bibinfo {author} {\bibfnamefont {D.}~\bibnamefont {Neilson}}, \
	and\ \bibinfo {author} {\bibfnamefont {A.~R.}\ \bibnamefont {Hamilton}},\
}\href {\doibase 10.1103/PhysRevLett.110.146803} {\bibfield  {journal}
{\bibinfo  {journal} {Phys. Rev. Lett.}\ }\textbf {\bibinfo {volume} {110}},\
\bibinfo {pages} {146803} (\bibinfo {year} {2013})}\BibitemShut {NoStop}%
\bibitem [{\citenamefont {Zarenia}\ \emph {et~al.}(2016)\citenamefont
	{Zarenia}, \citenamefont {Perali}, \citenamefont {Peeters},\ and\
	\citenamefont {Neilson}}]{Zarenia:nanoribbon16}%
\BibitemOpen
\bibfield  {author} {\bibinfo {author} {\bibfnamefont {M.}~\bibnamefont
		{Zarenia}}, \bibinfo {author} {\bibfnamefont {A.}~\bibnamefont {Perali}},
	\bibinfo {author} {\bibfnamefont {F.~M.}\ \bibnamefont {Peeters}}, \ and\
	\bibinfo {author} {\bibfnamefont {D.}~\bibnamefont {Neilson}},\ }\href
{\doibase 10.1038/srep24860} {\bibfield  {journal} {\bibinfo  {journal}
		{Scientific Reports}\ }\textbf {\bibinfo {volume} {6}},\ \bibinfo {pages}
	{24860} (\bibinfo {year} {2016})}\BibitemShut {NoStop}%
\bibitem [{\citenamefont {Zhang}\ and\ \citenamefont
	{Rossi}(2013)}]{PhysRevLett.111.086804}%
\BibitemOpen
\bibfield  {author} {\bibinfo {author} {\bibfnamefont {J.}~\bibnamefont
		{Zhang}}\ and\ \bibinfo {author} {\bibfnamefont {E.}~\bibnamefont {Rossi}},\
}\href {\doibase 10.1103/PhysRevLett.111.086804} {\bibfield  {journal}
{\bibinfo  {journal} {Phys. Rev. Lett.}\ }\textbf {\bibinfo {volume} {111}},\
\bibinfo {pages} {086804} (\bibinfo {year} {2013})}\BibitemShut {NoStop}%
\bibitem [{\citenamefont {Berman}\ and\ \citenamefont
	{Kezerashvili}(2016)}]{PhysRevB.93.245410}%
\BibitemOpen
\bibfield  {author} {\bibinfo {author} {\bibfnamefont {O.~L.}\ \bibnamefont
		{Berman}}\ and\ \bibinfo {author} {\bibfnamefont {R.~Y.}\ \bibnamefont
		{Kezerashvili}},\ }\href {\doibase 10.1103/PhysRevB.93.245410} {\bibfield
	{journal} {\bibinfo  {journal} {Phys. Rev. B}\ }\textbf {\bibinfo {volume}
		{93}},\ \bibinfo {pages} {245410} (\bibinfo {year} {2016})}\BibitemShut
{NoStop}%
\bibitem [{\citenamefont {Su}\ and\ \citenamefont
	{MacDonald}(2017)}]{PhysRevB.95.045416}%
\BibitemOpen
\bibfield  {author} {\bibinfo {author} {\bibfnamefont {J.-J.}\ \bibnamefont
		{Su}}\ and\ \bibinfo {author} {\bibfnamefont {A.~H.}\ \bibnamefont
		{MacDonald}},\ }\href {\doibase 10.1103/PhysRevB.95.045416} {\bibfield
	{journal} {\bibinfo  {journal} {Phys. Rev. B}\ }\textbf {\bibinfo {volume}
		{95}},\ \bibinfo {pages} {045416} (\bibinfo {year} {2017})}\BibitemShut
{NoStop}%
\bibitem [{\citenamefont {Li}\ \emph {et~al.}(2016)\citenamefont {Li},
	\citenamefont {Taniguchi}, \citenamefont {Watanabe}, \citenamefont {Hone},
	\citenamefont {Levchenko},\ and\ \citenamefont
	{Dean}}]{PhysRevLett.117.046802}%
\BibitemOpen
\bibfield  {author} {\bibinfo {author} {\bibfnamefont {J.~I.~A.}\
		\bibnamefont {Li}}, \bibinfo {author} {\bibfnamefont {T.}~\bibnamefont
		{Taniguchi}}, \bibinfo {author} {\bibfnamefont {K.}~\bibnamefont {Watanabe}},
	\bibinfo {author} {\bibfnamefont {J.}~\bibnamefont {Hone}}, \bibinfo {author}
	{\bibfnamefont {A.}~\bibnamefont {Levchenko}}, \ and\ \bibinfo {author}
	{\bibfnamefont {C.~R.}\ \bibnamefont {Dean}},\ }\href {\doibase
	10.1103/PhysRevLett.117.046802} {\bibfield  {journal} {\bibinfo  {journal}
		{Phys. Rev. Lett.}\ }\textbf {\bibinfo {volume} {117}},\ \bibinfo {pages}
	{046802} (\bibinfo {year} {2016})}\BibitemShut {NoStop}%
\bibitem [{\citenamefont {Joo}\ \emph {et~al.}(2017)\citenamefont {Joo},
	\citenamefont {Jin}, \citenamefont {Moon}, \citenamefont {Kim}, \citenamefont
	{Lee},\ and\ \citenamefont {Lee}}]{Joo2017}%
\BibitemOpen
\bibfield  {author} {\bibinfo {author} {\bibfnamefont {M.-K.}\ \bibnamefont
		{Joo}}, \bibinfo {author} {\bibfnamefont {Y.}~\bibnamefont {Jin}}, \bibinfo
	{author} {\bibfnamefont {B.~H.}\ \bibnamefont {Moon}}, \bibinfo {author}
	{\bibfnamefont {H.}~\bibnamefont {Kim}}, \bibinfo {author} {\bibfnamefont
		{S.}~\bibnamefont {Lee}}, \ and\ \bibinfo {author} {\bibfnamefont {Y.~H.}\
		\bibnamefont {Lee}},\ }\href {https://arxiv.org/abs/1711.00606} {\bibfield
	{journal} {\bibinfo  {journal} {arXiv:1711.00606}\ } (\bibinfo {year}
	{2017})}\BibitemShut {NoStop}%
\bibitem [{\citenamefont {Burg}\ \emph {et~al.}(2018)\citenamefont {Burg},
	\citenamefont {Prasad}, \citenamefont {Kim}, \citenamefont {Taniguchi},
	\citenamefont {Watanabe}, \citenamefont {MacDonald}, \citenamefont
	{Register},\ and\ \citenamefont {Tutuc}}]{GrapheneWSe2}%
\BibitemOpen
\bibfield  {author} {\bibinfo {author} {\bibfnamefont {G.~W.}\ \bibnamefont
		{Burg}}, \bibinfo {author} {\bibfnamefont {N.}~\bibnamefont {Prasad}},
	\bibinfo {author} {\bibfnamefont {K.}~\bibnamefont {Kim}}, \bibinfo {author}
	{\bibfnamefont {T.}~\bibnamefont {Taniguchi}}, \bibinfo {author}
	{\bibfnamefont {K.}~\bibnamefont {Watanabe}}, \bibinfo {author}
	{\bibfnamefont {A.~H.}\ \bibnamefont {MacDonald}}, \bibinfo {author}
	{\bibfnamefont {L.~F.}\ \bibnamefont {Register}}, \ and\ \bibinfo {author}
	{\bibfnamefont {E.}~\bibnamefont {Tutuc}},\ }\href
{https://arxiv.org/abs/1802.07331} {\bibfield  {journal} {\bibinfo  {journal}
		{arXiv:1802.07331}\ } (\bibinfo {year} {2018})}\BibitemShut {NoStop}%
\bibitem [{\citenamefont {Surrente}\ \emph {et~al.}(2016)\citenamefont
	{Surrente}, \citenamefont {Mitioglu}, \citenamefont {Galkowski},
	\citenamefont {Tabis}, \citenamefont {Maude},\ and\ \citenamefont
	{Plochocka}}]{PhysRevB.93.121405}%
\BibitemOpen
\bibfield  {author} {\bibinfo {author} {\bibfnamefont {A.}~\bibnamefont
		{Surrente}}, \bibinfo {author} {\bibfnamefont {A.~A.}\ \bibnamefont
		{Mitioglu}}, \bibinfo {author} {\bibfnamefont {K.}~\bibnamefont {Galkowski}},
	\bibinfo {author} {\bibfnamefont {W.}~\bibnamefont {Tabis}}, \bibinfo
	{author} {\bibfnamefont {D.~K.}\ \bibnamefont {Maude}}, \ and\ \bibinfo
	{author} {\bibfnamefont {P.}~\bibnamefont {Plochocka}},\ }\href {\doibase
	10.1103/PhysRevB.93.121405} {\bibfield  {journal} {\bibinfo  {journal} {Phys.
			Rev. B}\ }\textbf {\bibinfo {volume} {93}},\ \bibinfo {pages} {121405}
	(\bibinfo {year} {2016})}\BibitemShut {NoStop}%
\bibitem [{\citenamefont {Berman}\ \emph {et~al.}(2017)\citenamefont {Berman},
	\citenamefont {Gumbs},\ and\ \citenamefont
	{Kezerashvili}}]{PhysRevB.96.014505}%
\BibitemOpen
\bibfield  {author} {\bibinfo {author} {\bibfnamefont {O.~L.}\ \bibnamefont
		{Berman}}, \bibinfo {author} {\bibfnamefont {G.}~\bibnamefont {Gumbs}}, \
	and\ \bibinfo {author} {\bibfnamefont {R.~Y.}\ \bibnamefont {Kezerashvili}},\
}\href {\doibase 10.1103/PhysRevB.96.014505} {\bibfield  {journal} {\bibinfo
	{journal} {Phys. Rev. B}\ }\textbf {\bibinfo {volume} {96}},\ \bibinfo
{pages} {014505} (\bibinfo {year} {2017})}\BibitemShut {NoStop}%
\bibitem [{\citenamefont {Gortel}\ and\ \citenamefont
	{Swierkowski}(1996)}]{GORTEL1996146}%
\BibitemOpen
\bibfield  {author} {\bibinfo {author} {\bibfnamefont {Z.}~\bibnamefont
		{Gortel}}\ and\ \bibinfo {author} {\bibfnamefont {L.}~\bibnamefont
		{Swierkowski}},\ }\href {https://doi.org/10.1016/0039-6028(96)00355-X}
{\bibfield  {journal} {\bibinfo  {journal} {Surface Science}\ }\textbf
	{\bibinfo {volume} {361-362}},\ \bibinfo {pages} {146} (\bibinfo {year}
	{1996})}\BibitemShut {NoStop}%
\bibitem [{\citenamefont {Rodin}\ \emph {et~al.}(2014)\citenamefont {Rodin},
	\citenamefont {Carvalho},\ and\ \citenamefont {Castro~Neto}}]{Rodin:prl14}%
\BibitemOpen
\bibfield  {author} {\bibinfo {author} {\bibfnamefont {A.~S.}\ \bibnamefont
		{Rodin}}, \bibinfo {author} {\bibfnamefont {A.}~\bibnamefont {Carvalho}}, \
	and\ \bibinfo {author} {\bibfnamefont {A.~H.}\ \bibnamefont {Castro~Neto}},\
}\href {\doibase 10.1103/PhysRevLett.112.176801} {\bibfield  {journal}
{\bibinfo  {journal} {Phys. Rev. Lett.}\ }\textbf {\bibinfo {volume} {112}},\
\bibinfo {pages} {176801} (\bibinfo {year} {2014})}\BibitemShut {NoStop}%
\bibitem [{\citenamefont {Sodemann}\ \emph {et~al.}(2012)\citenamefont
	{Sodemann}, \citenamefont {Pesin},\ and\ \citenamefont
	{MacDonald}}]{PhysRevB.85.195136}%
\BibitemOpen
\bibfield  {author} {\bibinfo {author} {\bibfnamefont {I.}~\bibnamefont
		{Sodemann}}, \bibinfo {author} {\bibfnamefont {D.~A.}\ \bibnamefont {Pesin}},
	\ and\ \bibinfo {author} {\bibfnamefont {A.~H.}\ \bibnamefont {MacDonald}},\
}\href {\doibase 10.1103/PhysRevB.85.195136} {\bibfield  {journal} {\bibinfo
	{journal} {Phys. Rev. B}\ }\textbf {\bibinfo {volume} {85}},\ \bibinfo
{pages} {195136} (\bibinfo {year} {2012})}\BibitemShut {NoStop}%
\bibitem [{\citenamefont {Hwang}\ and\ \citenamefont
	{Das~Sarma}(2009)}]{Hwang:prb09}%
\BibitemOpen
\bibfield  {author} {\bibinfo {author} {\bibfnamefont {E.~H.}\ \bibnamefont
		{Hwang}}\ and\ \bibinfo {author} {\bibfnamefont {S.}~\bibnamefont
		{Das~Sarma}},\ }\href {\doibase 10.1103/PhysRevB.80.205405} {\bibfield
	{journal} {\bibinfo  {journal} {Phys. Rev. B}\ }\textbf {\bibinfo {volume}
		{80}},\ \bibinfo {pages} {205405} (\bibinfo {year} {2009})}\BibitemShut
{NoStop}%
\bibitem [{\citenamefont {Zhu}\ \emph {et~al.}(2013)\citenamefont {Zhu},
	\citenamefont {Badalyan},\ and\ \citenamefont
	{Peeters}}]{PhysRevB.87.085401}%
\BibitemOpen
\bibfield  {author} {\bibinfo {author} {\bibfnamefont {J.-J.}\ \bibnamefont
		{Zhu}}, \bibinfo {author} {\bibfnamefont {S.~M.}\ \bibnamefont {Badalyan}}, \
	and\ \bibinfo {author} {\bibfnamefont {F.~M.}\ \bibnamefont {Peeters}},\
}\href {\doibase 10.1103/PhysRevB.87.085401} {\bibfield  {journal} {\bibinfo
	{journal} {Phys. Rev. B}\ }\textbf {\bibinfo {volume} {87}},\ \bibinfo
{pages} {085401} (\bibinfo {year} {2013})}\BibitemShut {NoStop}%
\bibitem [{\citenamefont {Liu}\ \emph {et~al.}(2016)\citenamefont {Liu},
	\citenamefont {Low},\ and\ \citenamefont {Ruden}}]{PhysRevB.93.165402}%
\BibitemOpen
\bibfield  {author} {\bibinfo {author} {\bibfnamefont {Y.}~\bibnamefont
		{Liu}}, \bibinfo {author} {\bibfnamefont {T.}~\bibnamefont {Low}}, \ and\
	\bibinfo {author} {\bibfnamefont {P.~P.}\ \bibnamefont {Ruden}},\ }\href
{\doibase 10.1103/PhysRevB.93.165402} {\bibfield  {journal} {\bibinfo
		{journal} {Phys. Rev. B}\ }\textbf {\bibinfo {volume} {93}},\ \bibinfo
	{pages} {165402} (\bibinfo {year} {2016})}\BibitemShut {NoStop}%
\bibitem [{\citenamefont {Low}\ \emph {et~al.}(2014{\natexlab{b}})\citenamefont
	{Low}, \citenamefont {Rodin}, \citenamefont {Carvalho}, \citenamefont
	{Jiang}, \citenamefont {Wang}, \citenamefont {Xia},\ and\ \citenamefont
	{Castro~Neto}}]{Low:prb14}%
\BibitemOpen
\bibfield  {author} {\bibinfo {author} {\bibfnamefont {T.}~\bibnamefont
		{Low}}, \bibinfo {author} {\bibfnamefont {A.~S.}\ \bibnamefont {Rodin}},
	\bibinfo {author} {\bibfnamefont {A.}~\bibnamefont {Carvalho}}, \bibinfo
	{author} {\bibfnamefont {Y.}~\bibnamefont {Jiang}}, \bibinfo {author}
	{\bibfnamefont {H.}~\bibnamefont {Wang}}, \bibinfo {author} {\bibfnamefont
		{F.}~\bibnamefont {Xia}}, \ and\ \bibinfo {author} {\bibfnamefont {A.~H.}\
		\bibnamefont {Castro~Neto}},\ }\href {\doibase 10.1103/PhysRevB.90.075434}
{\bibfield  {journal} {\bibinfo  {journal} {Phys. Rev. B}\ }\textbf {\bibinfo
		{volume} {90}},\ \bibinfo {pages} {075434} (\bibinfo {year}
	{2014}{\natexlab{b}})}\BibitemShut {NoStop}%
\bibitem [{\citenamefont {Conti}\ \emph {et~al.}(2017)\citenamefont {Conti},
	\citenamefont {Perali}, \citenamefont {Peeters},\ and\ \citenamefont
	{Neilson}}]{PhysRevLett.119.257002}%
\BibitemOpen
\bibfield  {author} {\bibinfo {author} {\bibfnamefont {S.}~\bibnamefont
		{Conti}}, \bibinfo {author} {\bibfnamefont {A.}~\bibnamefont {Perali}},
	\bibinfo {author} {\bibfnamefont {F.~M.}\ \bibnamefont {Peeters}}, \ and\
	\bibinfo {author} {\bibfnamefont {D.}~\bibnamefont {Neilson}},\ }\href
{\doibase 10.1103/PhysRevLett.119.257002} {\bibfield  {journal} {\bibinfo
		{journal} {Phys. Rev. Lett.}\ }\textbf {\bibinfo {volume} {119}},\ \bibinfo
	{pages} {257002} (\bibinfo {year} {2017})}\BibitemShut {NoStop}%
\bibitem [{\citenamefont {Sun}\ \emph {et~al.}(2013)\citenamefont {Sun},
	\citenamefont {Zhu}, \citenamefont {Liu},\ and\ \citenamefont
	{Ji}}]{PhysRevA.88.063637}%
\BibitemOpen
\bibfield  {author} {\bibinfo {author} {\bibfnamefont {Q.}~\bibnamefont
		{Sun}}, \bibinfo {author} {\bibfnamefont {G.-B.}\ \bibnamefont {Zhu}},
	\bibinfo {author} {\bibfnamefont {W.-M.}\ \bibnamefont {Liu}}, \ and\
	\bibinfo {author} {\bibfnamefont {A.-C.}\ \bibnamefont {Ji}},\ }\href
{\doibase 10.1103/PhysRevA.88.063637} {\bibfield  {journal} {\bibinfo
		{journal} {Phys. Rev. A}\ }\textbf {\bibinfo {volume} {88}},\ \bibinfo
	{pages} {063637} (\bibinfo {year} {2013})}\BibitemShut {NoStop}%
\bibitem [{\citenamefont {Cao}\ \emph {et~al.}(2015)\citenamefont {Cao},
	\citenamefont {Liu}, \citenamefont {He}, \citenamefont {Long},\ and\
	\citenamefont {Hu}}]{PhysRevA.91.023609}%
\BibitemOpen
\bibfield  {author} {\bibinfo {author} {\bibfnamefont {Y.}~\bibnamefont
		{Cao}}, \bibinfo {author} {\bibfnamefont {X.-J.}\ \bibnamefont {Liu}},
	\bibinfo {author} {\bibfnamefont {L.}~\bibnamefont {He}}, \bibinfo {author}
	{\bibfnamefont {G.-L.}\ \bibnamefont {Long}}, \ and\ \bibinfo {author}
	{\bibfnamefont {H.}~\bibnamefont {Hu}},\ }\href {\doibase
	10.1103/PhysRevA.91.023609} {\bibfield  {journal} {\bibinfo  {journal} {Phys.
			Rev. A}\ }\textbf {\bibinfo {volume} {91}},\ \bibinfo {pages} {023609}
	(\bibinfo {year} {2015})}\BibitemShut {NoStop}%
\bibitem [{\citenamefont {Li}\ and\ \citenamefont
	{Appelbaum}(2014)}]{PhysRevB.90.115439}%
\BibitemOpen
\bibfield  {author} {\bibinfo {author} {\bibfnamefont {P.}~\bibnamefont
		{Li}}\ and\ \bibinfo {author} {\bibfnamefont {I.}~\bibnamefont {Appelbaum}},\
}\href {\doibase 10.1103/PhysRevB.90.115439} {\bibfield  {journal} {\bibinfo
	{journal} {Phys. Rev. B}\ }\textbf {\bibinfo {volume} {90}},\ \bibinfo
{pages} {115439} (\bibinfo {year} {2014})}\BibitemShut {NoStop}%
\bibitem [{\citenamefont {Rios}\ \emph {et~al.}()\citenamefont {Rios},
	\citenamefont {Perali}, \citenamefont {Needs},\ and\ \citenamefont
	{Neilson}}]{Pablo2017}%
\BibitemOpen
\bibfield  {author} {\bibinfo {author} {\bibfnamefont {P.~L.}\ \bibnamefont
		{Rios}}, \bibinfo {author} {\bibfnamefont {A.}~\bibnamefont {Perali}},
	\bibinfo {author} {\bibfnamefont {R.~J.}\ \bibnamefont {Needs}}, \ and\
	\bibinfo {author} {\bibfnamefont {D.}~\bibnamefont {Neilson}},\ }\href
{https://arxiv.org/abs/1710.06863} {\bibinfo  {journal} {arXiv:1710.06863,
		Physical Review Letters (2018), to appear.}\ }\BibitemShut {NoStop}%
\end{thebibliography}
\end{document}